\documentclass[conference]{IEEEtran}
\IEEEoverridecommandlockouts

\usepackage{cite}
\usepackage{amsmath,amssymb,amsfonts}
\usepackage{algorithmicx}
\usepackage{algpseudocode}
\usepackage{graphicx}
\usepackage{textcomp}
\usepackage[dvipsnames]{xcolor}

\usepackage{booktabs}
\usepackage{array}
\usepackage{multirow}

\usepackage{dirtree}
\usepackage{pifont}
\usepackage{tikz}

\usepackage{listings}
\usepackage{soul}

\usepackage{fancyhdr}

\usepackage{xcolor}

\usepackage[normalem]{ulem}

\usepackage[
    bookmarks=true,
    breaklinks=true,
    letterpaper=true,
    colorlinks,
    citecolor=blue,
    linkcolor=blue,
    urlcolor=blue
]{hyperref}

\usepackage[most]{tcolorbox}
\newtcolorbox{takeawaybox}{
    colback=black!5!white,     
    colframe=black!75!white,   
    boxrule=0.5pt,
    arc=1mm,
    left=4pt,
    right=4pt,
    top=5pt,
    bottom=5pt,
    boxsep=1pt,
}

\usepackage[
    caption=false,
    font=normalsize,
    labelfont=sf,
    textfont=sf
]{subfig}

\newcommand{\cmark}{\textcolor{green!60!black}{\large\ding{51}}}
\newcommand{\xmark}{\textcolor{red!60!red}{\large\ding{55}}}
\newcommand{\qmark}{\textcolor{orange!60!red}{\Large\textbf{?}}}

\def\fixme#1{\bgroup \color{red}{[{#1}]}\egroup}
\def\pr#1{\bgroup \color{blue}{[{#1}]}\egroup}
\def\remove#1{\bgroup \color{blue}{\st{#1}}\egroup}

\newcommand{\sixpt}{\fontsize{6.4pt}{7.2pt}\selectfont}
\DeclareRobustCommand{\circled}[1]{%
  \tikz[baseline=(myanchor.base)]
    \node[circle,fill=black,inner sep=.5pt] (myanchor)
    {\color{white}\footnotesize #1};%
}

\makeatletter
\renewcommand{\ALG@beginalgorithmic}{\scriptsize}
\makeatother

\definecolor{codegreen}{rgb}{0,0.6,0}
\definecolor{codegray}{rgb}{0.5,0.5,0.5}
\definecolor{codepurple}{rgb}{0.58,0,0.82}
\definecolor{backcolour}{rgb}{1,1,1}

\lstdefinestyle{mystyle}{
    backgroundcolor=\color{backcolour},
    commentstyle=\itshape\color{codegreen},
    keywordstyle=\color{magenta},
    numberstyle=\tiny\color{codegray},
    stringstyle=\color{codepurple},
    basicstyle=\scriptsize,
    breakatwhitespace=false,
    breaklines=true,
    captionpos=b,
    keepspaces=true,
    numbers=left,
    numbersep=5pt,
    showspaces=false,
    showstringspaces=false,
    showtabs=false,
    tabsize=7,
    frame=single
}

\newsavebox{\codebaseline}
\newsavebox{\codeupdated}

\begin{document}


\fancypagestyle{firstpage}{
  \fancyhf{}
  \renewcommand{\headrulewidth}{0pt}

  \fancyhead[C]{%
    \scriptsize
    This article has been accepted for publication in IEEE International Symposium on Workload Characterization (IISWC), 2026 (author version).
  }

  \fancyfoot[C]{\thepage}
}

\title{
Integration, Enhancements and Evaluation of \\ Memory Simulators
}


\author{%
  \IEEEauthorblockN{%
    \IEEEauthorrefmark{1}Pouya~Esmaili-Dokht\IEEEauthorrefmark{2}\IEEEauthorrefmark{4},
    \IEEEauthorrefmark{1}Arash~Yadegari\IEEEauthorrefmark{4}\IEEEauthorrefmark{3},
    Victor~Xirau\IEEEauthorrefmark{4},
    Julian~Pavon\IEEEauthorrefmark{4},
    Hamid~Sarbazi-Azad\IEEEauthorrefmark{3}\IEEEauthorrefmark{6},\\
    Adri\'an~Cristal\IEEEauthorrefmark{2}\IEEEauthorrefmark{4},
    Eduard~Ayguad\'e\IEEEauthorrefmark{2}\IEEEauthorrefmark{4},
    Petar~Radojkovi\'c\IEEEauthorrefmark{4}
  }

  \IEEEauthorblockA{%
    Universitat Polit\`ecnica de Catalunya\IEEEauthorrefmark{2},
    Barcelona Supercomputing Center\IEEEauthorrefmark{4},
    Sharif University of Technology\IEEEauthorrefmark{3},
    IPM\IEEEauthorrefmark{6}
  }

  \IEEEauthorblockA{%
    \{pouya.esmaili,
    arash.yadegari,
    victor.xirau,
    julian.pavon,
    adrian.cristal,
    eduard.ayguade,
    petar.radojkovic\}@bsc.es
  }
  \IEEEauthorblockA{%
    azad@sharif.edu
  }

  \thanks{\IEEEauthorrefmark{1}Both authors contributed equally to this study.}
}


\maketitle
\thispagestyle{firstpage}
\pagestyle{plain}


\begin{abstract}

Currently, there is no common methodology or set of practices for evaluating memory-system performance in integrated simulation infrastructures. As a result, studies adopt different evaluation approaches, which can lead to conflicting conclusions even for the same simulation infrastructure.

We propose a validation methodology that evaluates memory-system performance from three complementary views: the memory simulator, the CPU--memory interface and the simulated application. We use these views to detect, understand and correct memory-simulation inaccuracies arising from simulator integration and simplified modeling of microarchitectural features. The methodology reveals that the application's perception of memory performance can become effectively decoupled from the detailed memory simulation. Most importantly, we identify and correct a fundamental memory-model mismatch between delayed-response cycle-accurate memory simulators and ZSim, a representative parallel CPU simulator.

We demonstrate the broader applicability of these enhancements by applying them to ZSim integrations with Ramulator\,2, DRAMsim3 and DRAMSys\,4, covering all major public cycle-accurate memory simulators. After the corrections and enhancements, all integrations reproduce the main bandwidth--latency trends and key characteristics of the actual system. Across the evaluated simulators, STREAM bandwidth differs from actual-hardware measurements by less than 5\% in most experiments, while pointer-chase latency remains 15--18\% lower than on actual hardware. All simulator enhancements are released as open source.





\end{abstract}




\section{Background and introduction}
\label{sec:Introduction}

\looseness -1 Memory-system performance models are a core component of hardware simulation infrastructures used in computer-architecture research. Based on when they return a memory-access latency estimate, these models can be broadly divided into two categories: immediate-response models and delayed-response cycle-accurate memory simulators.

\looseness -1 \textbf{Immediate-response models} provide a memory-access latency estimate with a single function call when the memory access is issued~\cite{Eyerman:IRMM}. Examples include fixed-latency models~\cite{Ganesh:queueModel}, queuing models~\cite{Gulur:memModel3queue}, and models based on hardware measurements~\cite{esmaili:mess,crowthers:panem}. These fast models are used by many CPU simulators, as shown in Table~\ref{tab:mem-models-transposed}. However, their memory-system abstractions may lack the level of detail required for design-space explorations of novel DRAM architectures and timings.

This level of detail is modeled by \textbf{cycle-accurate memory simulators}, such as 
DRAMsim3~\cite{Shangli:dramsim3}, Ramulator\,2~\cite{Haocong:Ramulator2} and DRAMSys\,4~\cite{steiner:dramsys4}.   
 These simulators determine memory-access latency by modeling memory-device hardware structures and timings, as well as scheduling, ordering and contention among concurrent memory requests~\cite{kim:ramulator,Rosenfeld:dramsim2,Wang:DRAMSIM,chatterjee:usimm}.  
The downside is that cycle-accurate memory simulators cannot provide an immediate latency response when a request is issued. 
This makes them incompatible with some parallel simulation infrastructures that are essential for simulating large and complex state-of-the-art CPUs (see Table\,\ref{tab:mem-models-transposed}). As a result, simulator users often have to choose between cycle-accurate memory models and fast parallel system simulators~\cite{Eyerman:IRMM,Shang:misscycle}.

\begin{table}[t!]
\centering
\caption{Simulator users frequently have to choose between cycle-accurate memory models and fast parallel system simulators.} 
\label{tab:mem-models-transposed}

\renewcommand{\arraystretch}{1.2}

\resizebox{\linewidth}{!}{%
\begin{tabular}{@{}l l
>{\centering\arraybackslash}m{1cm}
>{\centering\arraybackslash}m{1cm}
>{\centering\arraybackslash}m{1cm}@{}}
\toprule
Category & Model & gem5 & Sniper & ZSim\\
\midrule
\multirow{3}{*}{\shortstack{\textbf{Immediate-response}}}
& Fixed-latency
& \cmark & \cmark & \cmark \\

& \shortstack{Queuing models (M/D/1)}
& \cmark & \cmark & \cmark \\

& \shortstack{Measurement-based (Mess)}
& \cmark & \cmark & \cmark \\

\midrule

\multirow{2}{*}{\shortstack{\textbf{Cycle-accurate } \\ \textbf{Delayed-response}}}
& DRAMsim    & \cmark & \xmark & \qmark \\
& Ramulator   & \cmark & \xmark & \qmark \\

\bottomrule
\end{tabular}%
}
\end{table}

ZSim~\cite{sanchez:zsim} is the only widely used parallel simulator that has been integrated with DRAMsim3~\cite{Shangli:dramsim3} and Ramulator~\cite{Oliveira:damov-ZSim-ramulator}.\footnote{Before our study, ZSim had not been integrated with Ramulator\,2.} However, recent studies have reached conflicting conclusions about the accuracy of these integrations~\cite{bostanci:ramulator-damov-safari, esmaili:mess, Shang:misscycle}. The DRAMsim3 developers reported fundamental memory-model incompatibilities in their ZSim integration~\cite{Shang:misscycle}, and the corresponding integration is currently unavailable~\cite{ZSimDRAMsim3:repo}. A subsequent study reported significant inaccuracies in both ZSim+DRAMsim3 and ZSim+Ramulator and broadly attributed them to the CPU--memory interfaces~\cite{esmaili:mess}. In contrast, a more recent study concluded that Ramulator integrated with ZSim through the DAMOV interface accurately models memory-system performance~\cite{bostanci:ramulator-damov-safari}.

Our study shows that these conflicting conclusions largely arise from differences in how simulated memory performance is evaluated. This is not surprising, as the literature provides no common methodology for validating memory-system performance in simulation infrastructures.

Our study has three main contributions. First, we propose a methodology for evaluating memory models in integrated CPU--memory simulation infrastructures. The methodology examines memory-system performance from three complementary views: the \textbf{memory simulator}, the \textbf{CPU--memory interface} and the \textbf{simulated application}~(Section~\ref{sec:Methodology}). When applied to ZSim+Ramulator, it exposes large discrepancies between the statistics reported by the \textbf{memory simulator} and the memory performance perceived by the \textbf{simulated application}~(Section~\ref{sec:Init-validation}).
We further explain why these discrepancies can \textbf{remain undetected}, creating a potential \textbf{blind spot} for simulator developers and users. This demonstrates the risk of incomplete validation methodologies\,---\,they may \textbf{fail to expose substantial simulation inaccuracies} while still providing \textbf{apparent confidence} in the accuracy of the integrated simulation infrastructure.

Second, we use multiple memory-performance views to \textbf{detect}, \textbf{understand} and \textbf{correct} various memory-simulation inaccuracies arising from simulator integration (Section~\ref{sec:Integration}) and simplified modeling of some microarchitectural features (Section~\ref{sec:Further-enhance}).
Most importantly, we identify and correct a fundamental \textbf{memory-model mismatch} between \textbf{delayed-response cycle-accurate memory simulators} and ZSim, which is representative of a class of \textbf{parallel CPU simulators}.

This mismatch \textbf{decouples} the application's perception of main-memory performance from the rest of the memory-system simulation. As a result, although the memory simulator broadly follows the actual-system performance trend, the simulated load-to-use latency remains unrealistically low, starting at only 25\,ns.
While previous studies argue that parallel CPU simulators should use immediate-response memory models~\cite{Eyerman:IRMM, Shang:misscycle}, we show that the mismatch with delayed-response memory simulators can also be addressed through enhancements to the CPU--memory simulator interface. This demonstrates that cycle-accurate memory simulators can be integrated into massively parallel simulation infrastructures, motivating further work in this direction.


\looseness -1
Third, we demonstrate the broader applicability of these enhancements by applying them to \textbf{ZSim integrations} with \textbf{Ramulator\,2, DRAMsim3} and \textbf{DRAMSys\,4}, covering all major public cycle-accurate memory simulators (Section~\ref{sec:Quantitative-validation}). After these corrections, all integrations produce bandwidth--latency curves whose shapes and key characteristics closely resemble those of the actual system. They show slightly lower STREAM bandwidths, with differences below 5\%  in most experiments, and 15--18\% lower pointer-chase latencies than the actual hardware measurements.
All simulator enhancements are released as open source~\cite{esmaili:interfaceGit}.

In Section~\ref{sec:Effort}, we analyze the cost of applying the proposed validation methodology. We show that simple \textbf{sanity checks} can be performed at \textbf{negligible cost}, while more \textbf{detailed and holistic} memory-system evaluations can be completed in as little as \textbf{one hour of simulation time}.

\section{Experimental environment}
\label{sec:Methodology}

\subsection{Simulation platform} 

We use the ZSim CPU simulator\,\cite{sanchez:zsim, esmaili:acm} configured to model the pipeline and cache hierarchy of an Intel\,Xeon\,Platinum\,8160 processor\,\cite{intel_xeon_8160, nasa:skylake}. Table\,\ref{table:configuration} lists the main simulation parameters, which match the hardware features of the actual Intel server used to validate simulated 
\mbox{memory-system performance\,\cite{BSC:tech,intel_xeon_8160,samsung2017ddr4rdimm}.}

We start our analysis with ZSim integrated with Ramulator through the interface provided by the DAMOV simulation framework~\cite{Oliveira:damov-ZSim-ramulator}. DAMOV has been used to simulate both conventional CPU--memory systems and processing-in-memory architectures~\cite{Oliveira:damov-ZSim-ramulator, olivera:thesisdamov,  Ghiasi:ALPDamov}.

\looseness -1 To demonstrate the broader applicability of our work, Section\,\ref{sec:finalChart} also enhances and validates the memory performance of the integrated ZSim+DRAMsim3 infrastructure~\cite{esmaili:acm}. Finally, we integrate and validate ZSim with Ramulator~2~\cite{Haocong:Ramulator2} and DRAMSys~4~\cite{steiner:dramsys4}, which were not previously connected to ZSim, covering all major public cycle-accurate memory simulators.

\begin{table}[!t]
\centering
\caption{Simulation parameters configured to match the hardware features of the Intel Xeon Platinum 8160 validation system~\cite{BSC:tech,intel_xeon_8160,samsung2017ddr4rdimm}}
\label{table:configuration}
\begin{tabular*}{\linewidth}{@{\extracolsep{\fill}}ll}
\toprule
\textbf{ZSim: CPU parameters}   & \textbf{Ramulator: Memory  parameters}     \\
\midrule
Skylake microarchitecture       &	6$\times$ DDR4-2666~channels		     \\
24\,cores @ 2.1\,GHz          	&	1$\times$ 8\,GB DIMM per channel	  	 \\
~~~Private 32\,KB L1-D cache    &	2 ranks per DIMM 	                     \\
~~~Private 32\,KB L1-I cache    &	8 devices per rank                       \\
~~~Private 1\,MB L2 cache    	&	16 banks per device                      \\
Shared 33\,MB LLC cache     	&   32\,MB per device bank	                 \\

\bottomrule

\end{tabular*}
\end{table}

\subsection{Proposed validation methodology:  Multiple memory-performance views} 
\label{sec:3-view}

\begin{figure}[!t]
 \centering
 \includegraphics[width=\linewidth]{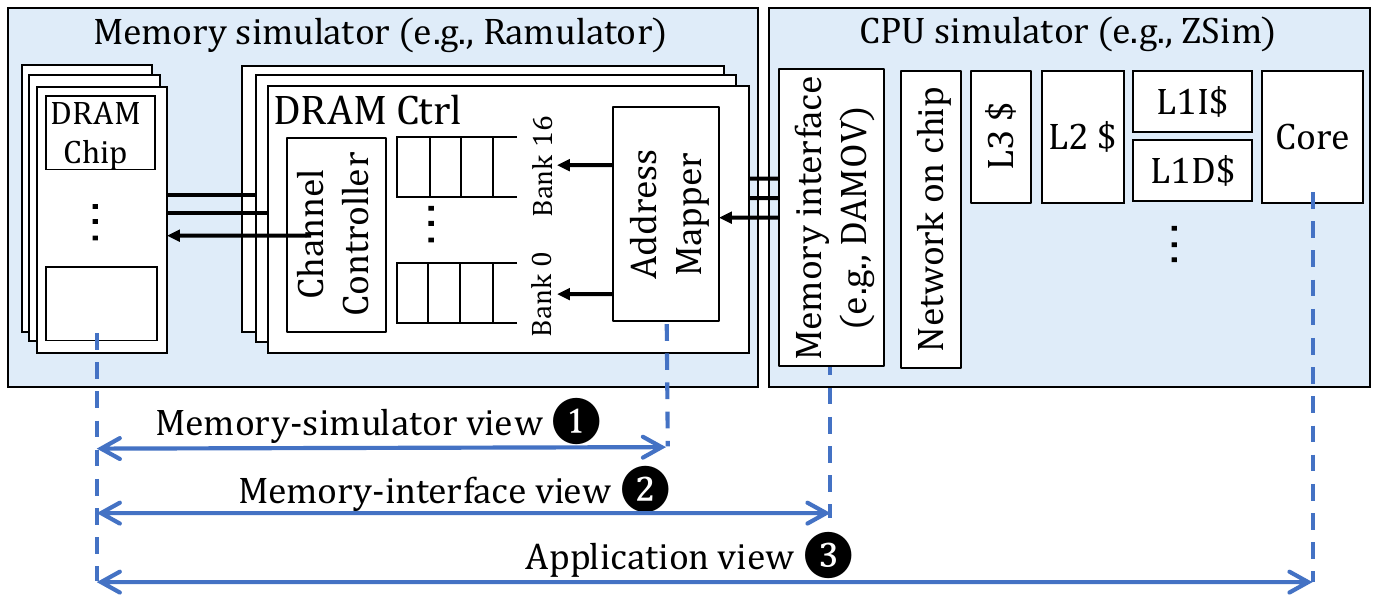}%
 \caption{Memory-simulator\,\circled{1} and memory-interface\,\circled{2} views expose internal simulator statistics. The application view\,\circled{3} is the final memory-simulation outcome that determines simulated application performance, making it the primary measure of memory-simulation accuracy.}
 \label{fig:3view-methodology}
\end{figure}

In integrated CPU--memory simulation platforms, memory-system performance can be evaluated at multiple points in the simulated execution pipeline. Figure\,\ref{fig:3view-methodology} shows the three memory-performance views considered in this study. The \textbf{memory-simulator view}\,\circled{1} uses statistics reported by the memory simulator, such as
DRAMsim~\cite{Shangli:dramsim3}, Ramulator~\cite{kim:ramulator, Haocong:Ramulator2} or DRAMSys~\cite{steiner:dramsys4}. 
This captures how the memory simulator processes requests received from the host simulator. 
The \textbf{memory-interface view}\,\circled{2} 
captures how the CPU simulator observes simulated memory performance 
and uses statistics collected at the CPU simulator's memory interface.  
Although the memory-simulator and memory-interface views are adjacent in the simulation platform, they can report substantially different performance. 
Finally, the \textbf{application view}\,\circled{3} measures the latency observed by a CPU core, i.e., the application's load-to-use latency for a main-memory access.

All three views are useful, but they serve different purposes. 
The memory-simulator and memory-interface views allow users to track memory-system performance across different stages of the integrated simulation infrastructure, which is essential for
\textbf{performance validation} and \textbf{debugging}~(Sections~\ref{sec:Init-validation} and~\ref{sec:Integration}). 
The application view directly determines simulated application performance and is therefore the relevant measure of \textbf{memory-simulation accuracy}.

\subsection{Benchmarks and methodology} 
Our memory-system performance characterization is based on the recently released Mess benchmark~\cite{esmaili:mess}. Mess characterizes memory-system behavior by measuring memory-access latency across a wide range of traffic intensities and read/write compositions. The resulting measurements are presented as families of bandwidth--latency curves.

\looseness -1  Mess measures memory-access latency using a pointer-chase benchmark running on one CPU core. When executed alone, pointer-chase measures unloaded memory-access latency. To measure loaded memory latency, Mess concurrently runs a configurable memory-traffic generator on the remaining CPU cores. 
Application-view latency is computed from the overall simulation time reported by ZSim.  
For the memory-simulator and memory-interface views, we measure the round-trip latency of pointer-chase requests from the corresponding simulated hardware components. 
We also account for latency penalties introduced by OS features that are present in the real system but not simulated by ZSim. Specifically, Mess data structures are allocated in huge pages to minimize the overhead of translation lookaside buffer misses and page walks on the real server. At runtime, these overheads are monitored with hardware counters and subtracted from the memory-latency measurements.

Mess reports memory bandwidth at the CPU--memory interface. In the real system, this bandwidth is measured using memory-controller hardware counters. In simulation, we compute the corresponding bandwidth from the number of memory requests exchanged between the ZSim memory interface and the external memory simulator.  
This bandwidth is reported with Ramulator's \texttt{ramulator.read\_bandwidth} and \texttt{ramulator.write\_bandwidth} statistics~\cite{bostanci:ramulator-damov-safari, kim:ramulator}. 

The Mess benchmark generates complex memory-traffic patterns determined by sequential accesses within each core and by the interleaving of memory requests across cores~\cite{esmaili:mess}. To confirm that our findings are not specific to these access patterns, we repeated all experiments using Mess-Random, a variant that generates random memory traffic on each core. The results led to the same conclusions.

In addition to the detailed Mess characterization, we compare real and simulated memory-system performance using two standard memory benchmarks: \textbf{pointer-chase}~\cite{Verdejo:microbenchmarks,sanchez:zsimEnhance} 
and \textbf{STREAM}~\cite{mccalpin:streamLink, mccalpin:streamBenchmark}. 
In these experiments, we focus on the application view, because it ultimately determines simulated benchmark performance. Standalone pointer-chase measurements follow the same methodology as the Mess latency measurements. STREAM reports the maximum sustained bandwidth of four kernels: 
\emph{Copy}, \emph{Scale}, \emph{Add} and \emph{Triad}. The benchmark computes sustained bandwidth from the size of the data structures, the number of load and store operations, and the actual or simulated execution time.\footnote{
STREAM accounts for application-level data traffic, whereas Mess measures architecture-level memory bandwidth using hardware counters. The latter includes both application data traffic and microarchitectural data transfers, e.g., those caused by write-allocate cache policies. The differences between Mess and STREAM bandwidth measurements are detailed in the Mess paper~\cite{esmaili:mess}.}

\section{Initial evaluation}
\label{sec:Init-validation}

\begin{figure*}[!t]
  \centering
  \includegraphics[width=.4\linewidth]{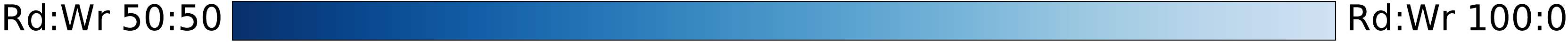}
\vspace{-.3cm}

  \subfloat[Intel\,Skylake\,CPU with 6$\times$DDR4-2666%
           \label{fig:actual-hw}]{%
    \includegraphics[width=.25\linewidth]{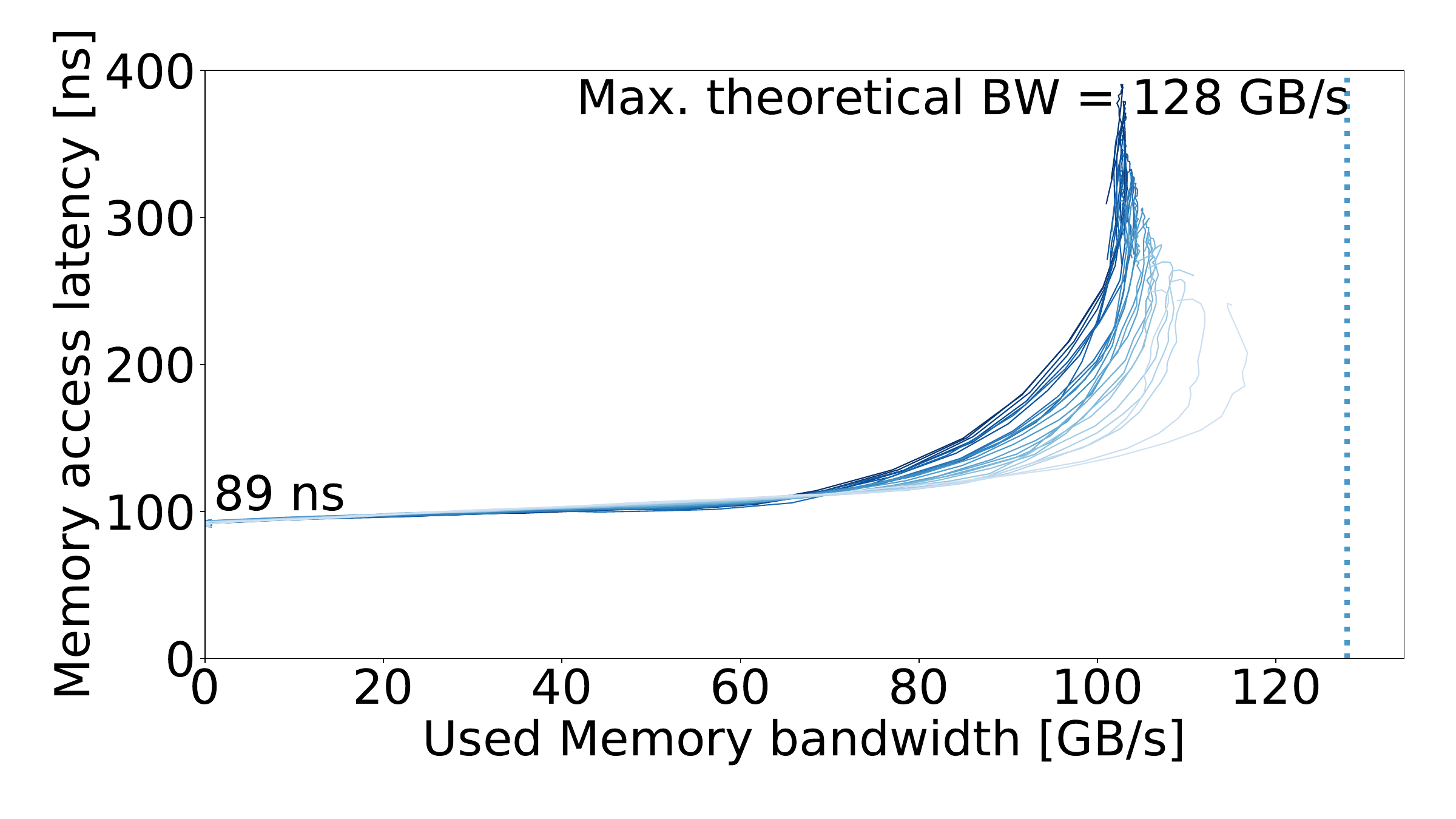}%
  }%
  \hfill
  \subfloat[Memory-simulator view%
           \label{fig:baseNoFix_ramulator}]{%
    \includegraphics[width=.25\linewidth]{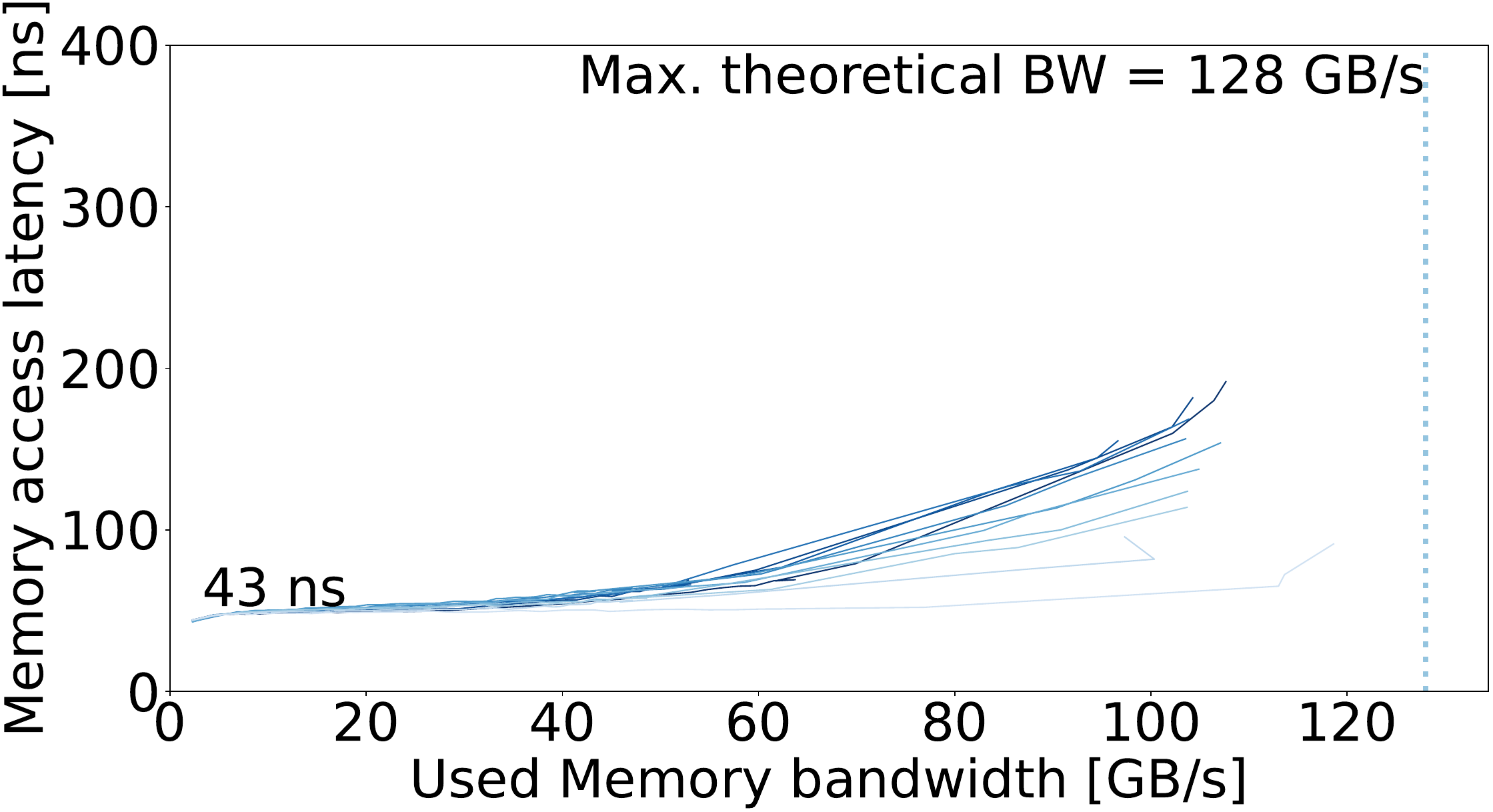}%
  }%
  \subfloat[Memory-interface view%
           \label{fig:baseNoFix_interface}]{%
    \includegraphics[width=.25\linewidth]{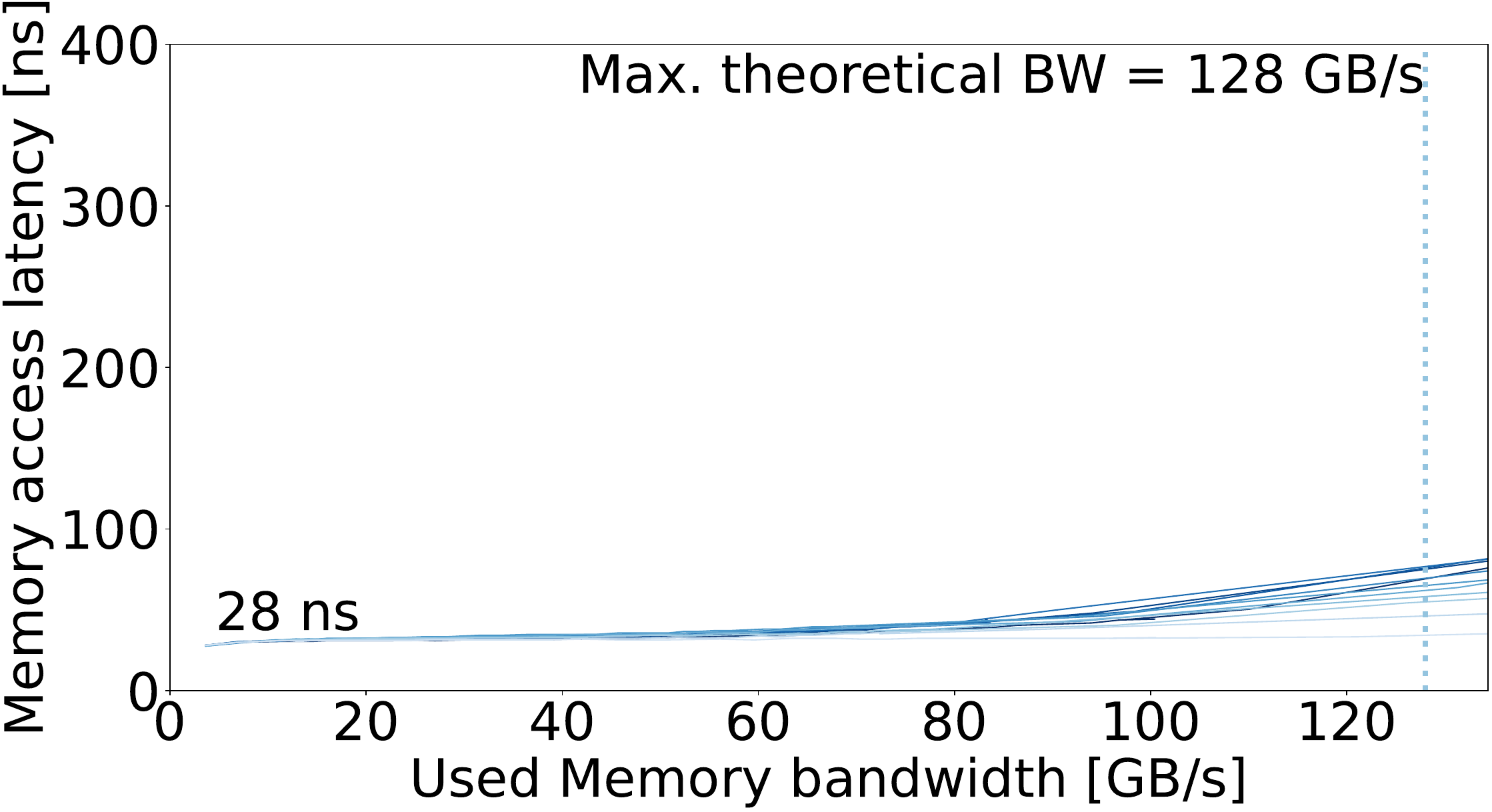}%
  }%
  \hfill
  \subfloat[Application view%
           \label{fig:baseNoFix_core}]{%
    \includegraphics[width=.25\linewidth]{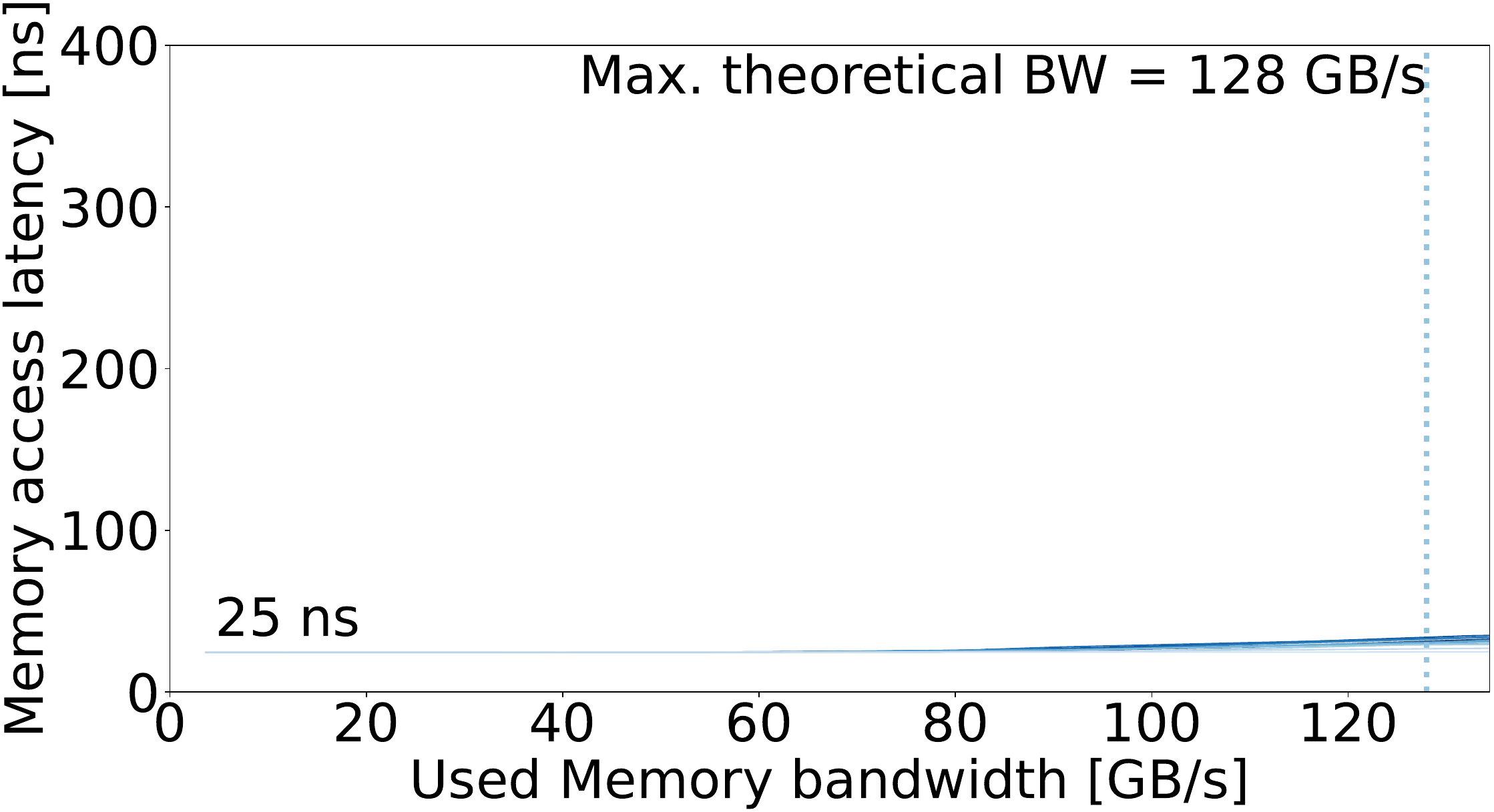}%
  }%
  \caption[Four subfigures]{%
    The application view of memory-access latency starts at only 25\,ns and remains nearly constant across the entire realistic bandwidth range~(Figure\,\ref{fig:baseNoFix_core}). 
    This behavior is decoupled from the actual system performance (Figure\,\ref{fig:actual-hw}) and, more surprisingly, from the memory simulator itself (Figure\,\ref{fig:baseNoFix_ramulator}).}  
  \label{fig:original-damov-interface}
\end{figure*}

Figure\,\ref{fig:original-damov-interface} compares the memory-system performance of the real Intel Xeon Platinum 8160 server with the three simulated performance views. The charts show Mess bandwidth--latency curves across the full range of memory-traffic intensity, from an unloaded to a fully saturated memory system. Different shades of blue represent different read/write compositions.


\subsection{Multiple memory-performance views} 

The \textbf{memory-simulator} view (Figure\,\ref{fig:baseNoFix_ramulator}), 
with some variations, broadly follows the actual-system performance trend.  
The simulated bandwidth utilization is similar to that of the actual system, 
with memory bandwidth saturating between 100\,GB/s and 120\,GB/s. 
The unloaded memory latency is 43\,ns, which is significantly below the 89\,ns measured in the actual system. 
Still, this difference is expected in part because the actual-system measurements capture the full load-to-use latency, 
whereas the memory-simulator view reports round-trip latency measured from the memory controller.  
Another difference appears in the saturated memory system (100--120\,GB/s), where the maximum simulated latency ranges from 90\,ns to 195\,ns, significantly below the 240--390\,ns measured in the actual system.  
Finally, the simulator correctly models the performance impact of read and write memory traffic. 
As in the actual hardware measurements, 100\%-read traffic achieves the best performance, with the lowest latency and the highest bandwidth, while memory writes reduce performance and cause the memory system to reach saturation sooner.

The \textbf{memory-interface} view (Figure\,\ref{fig:baseNoFix_interface}) shows two surprising results. 
First, memory latency starts at only 28\,ns and remains notably below the memory-simulator view across the entire bandwidth range. 
Second, the simulated memory bandwidth exceeds the theoretical maximum by 40\%.


Finally, the \textbf{application view} (Figure\,\ref{fig:baseNoFix_core}) shows an even larger discrepancy from actual memory-system performance. The maximum simulated bandwidth still significantly exceeds the theoretical maximum, while the simulated load-to-use latency starts at only 25\,ns and remains nearly constant across the entire realistic bandwidth range.
As expected, the STREAM and pointer-chase benchmark results are consistent with the Mess characterization. The simulated STREAM bandwidths significantly exceed the hardware measurements, while the pointer-chase results closely match the unloaded memory-access latency observed in the application view.
 
\looseness -1 This is an important issue for this integrated simulation infrastructure, because the application's memory-performance perception\,(Figure\,\ref{fig:baseNoFix_core}) is decoupled from the memory-simulator statistics\,(Figure\,\ref{fig:baseNoFix_ramulator}).  
To ensure that the inaccuracies are not caused by improper use of the DAMOV interface, we repeat the evaluation on the original DAMOV framework\,\cite{Oliveira:damov-ZSim-ramulator} and detect the same issue. 
The evaluation charts for the original DAMOV framework are presented in Appendix~\ref{sec:damov-evaluation}
and can be reproduced using the paper's repository~\cite{esmaili:interfaceGit}.

\begin{takeawaybox}
\textbf{Finding~1.}
In integrated CPU–memory simulation infrastructures, the application’s perception of memory
performance can become effectively decoupled from the detailed memory simulation. 
\end{takeawaybox}

\section{\mbox{ZSim+Ramulator integration}}
\label{sec:Integration}

\subsection{Problem\,1: Time-domain consistency} 
\label{sec:Time-domain-consistency}


Significant discrepancies between the memory-simulator view (Figure\,\ref{fig:baseNoFix_ramulator}) and the memory-interface view (Figure\,\ref{fig:baseNoFix_interface}) indicate that the CPU--memory interface introduces inaccuracies when handling memory requests. 

Inspection of the interface implementation revealed that one source of inaccuracy was the DAMOV interface block responsible for maintaining time-domain consistency between the CPU and memory simulators. 
Because memory operates at a lower frequency (1.333\,GHz) than the CPU (2.1\,GHz), its simulated cycle time, or tick, should be proportionally 1.57$\times$ longer.  
In the current simulator release~\cite{safari:damov-sim}, the block responsible for cross-simulator clocking was disabled, causing the CPU simulator to perceive the memory system as running at a 1.57$\times$ higher 
frequency.\footnote{ 
A comment in the DAMOV interface source code~\cite{safari:damov-interface-code} explicitly states that the cross-simulator clocking code should be enabled when the memory and host clocks differ. However, relying on users to manually uncomment this code is not a robust software-development practice~\cite{taschuk2017ten,wilson2017good,misra2023cpp}, as the required configuration can easily be overlooked. Indeed, this configuration was missed in two recent studies that evaluated the platform~\cite{esmaili:mess, bostanci:ramulator-damov-safari}, one of which was co-authored by the DAMOV developers~\cite{safari:ramulator-gate-artifact}.}  
This led to unrealistically high memory bandwidth on the CPU side of the memory interface, which was further propagated to the application view (Figure\,\ref{fig:baseNoFix_core}).

\begin{figure}[!t]
  \centering
  \includegraphics[width=.8\linewidth]{graphics/horizontal-legend}
\vspace{-.3cm}

  \subfloat[Memory-interface view%
           \label{fig:baseline-slightly-modify-interface}]{%
    \includegraphics[width=.5\linewidth]{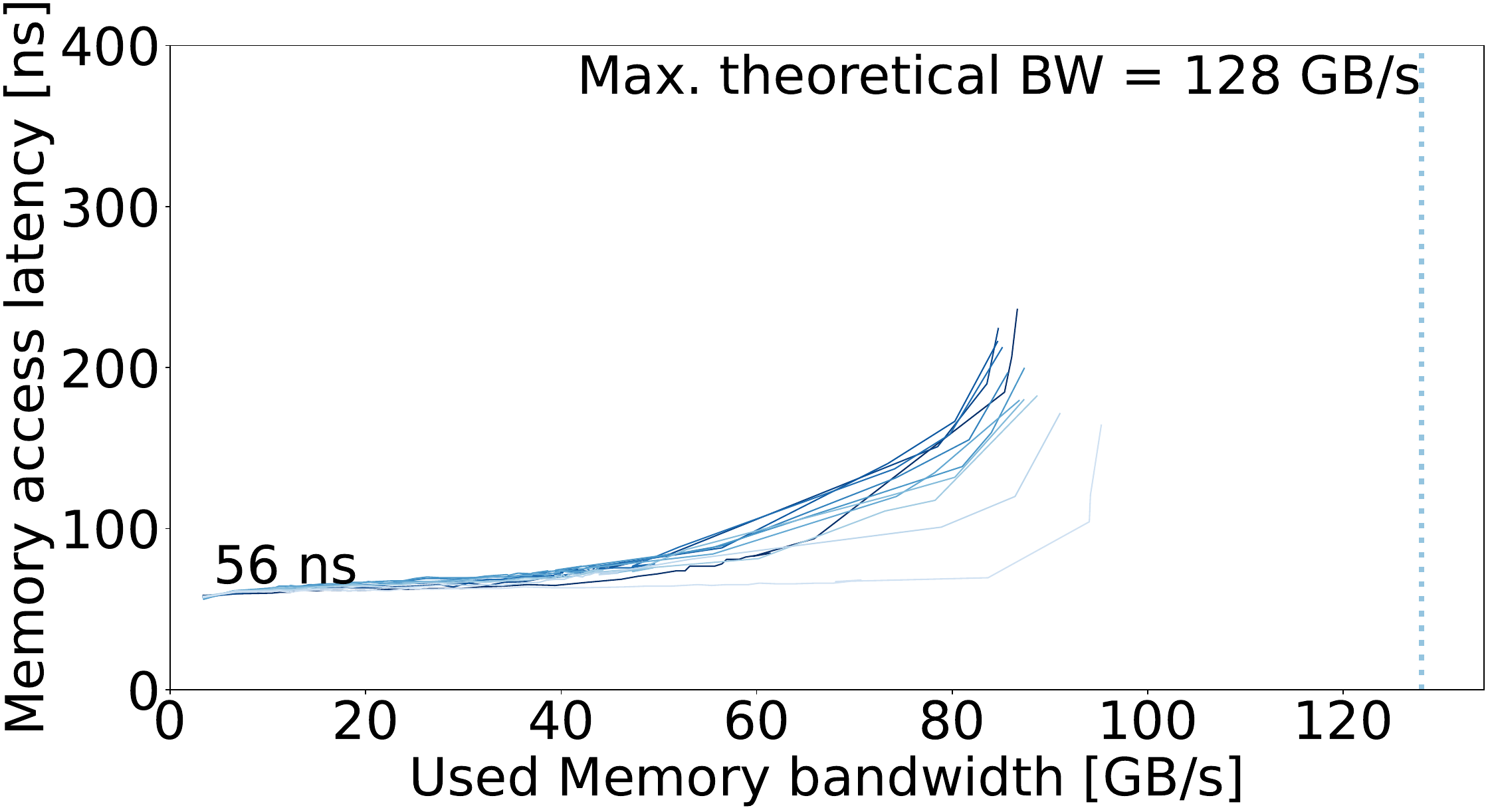}%
  }%
  \subfloat[Application view%
           \label{fig:baseline-slightly-modify-core}]{%
    \includegraphics[width=.5\linewidth]{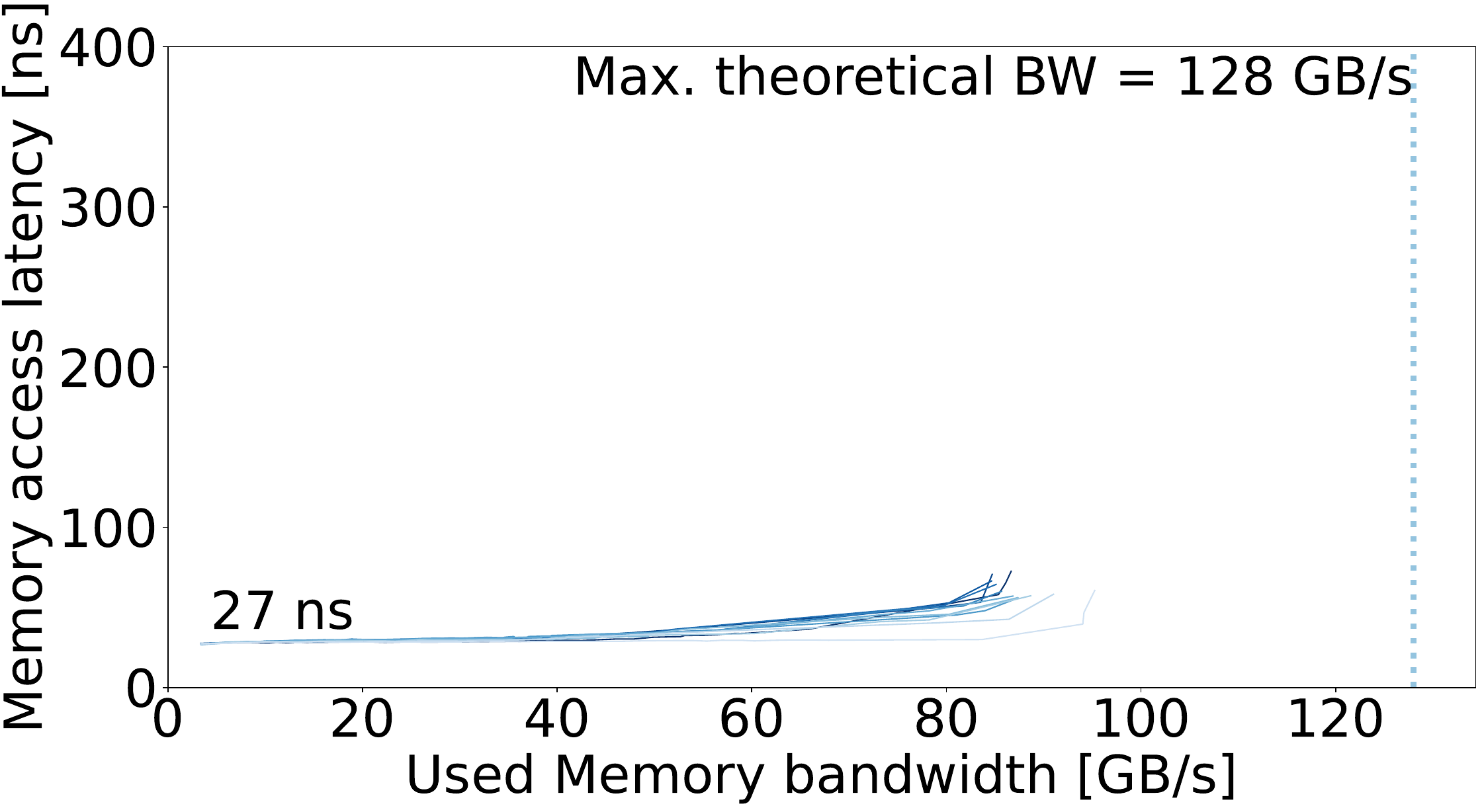}%
  }
  
  
  
  \caption[Four subfigures]{%
  Clock scaling in the memory-simulator interface corrects the unrealistically high bandwidths previously observed in the memory-interface and application views.} 
  \label{fig:baseline-slightly-modify}
\end{figure}

\begin{figure}[!t]
  \centering
\includegraphics[width=.8\linewidth]{graphics/horizontal-legend}
\vspace{-.3cm}

  \subfloat[Memory-interface view%
           \label{fig:correct-tick-interface}]{%
    \includegraphics[width=.5\linewidth]{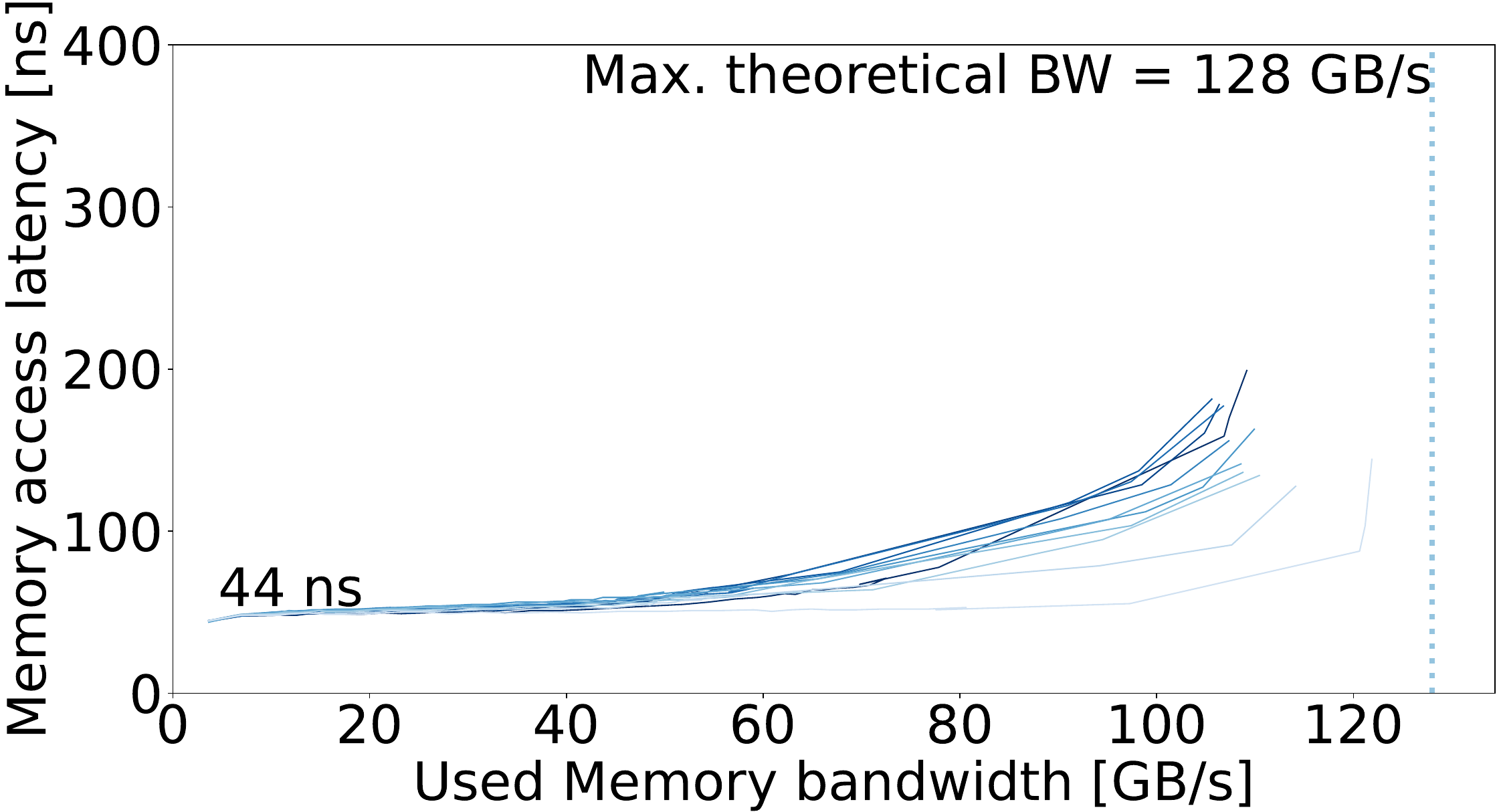}%
  }%
  \subfloat[Application view%
           \label{fig:correct-tick-core}]{%
    \includegraphics[width=.5\linewidth]{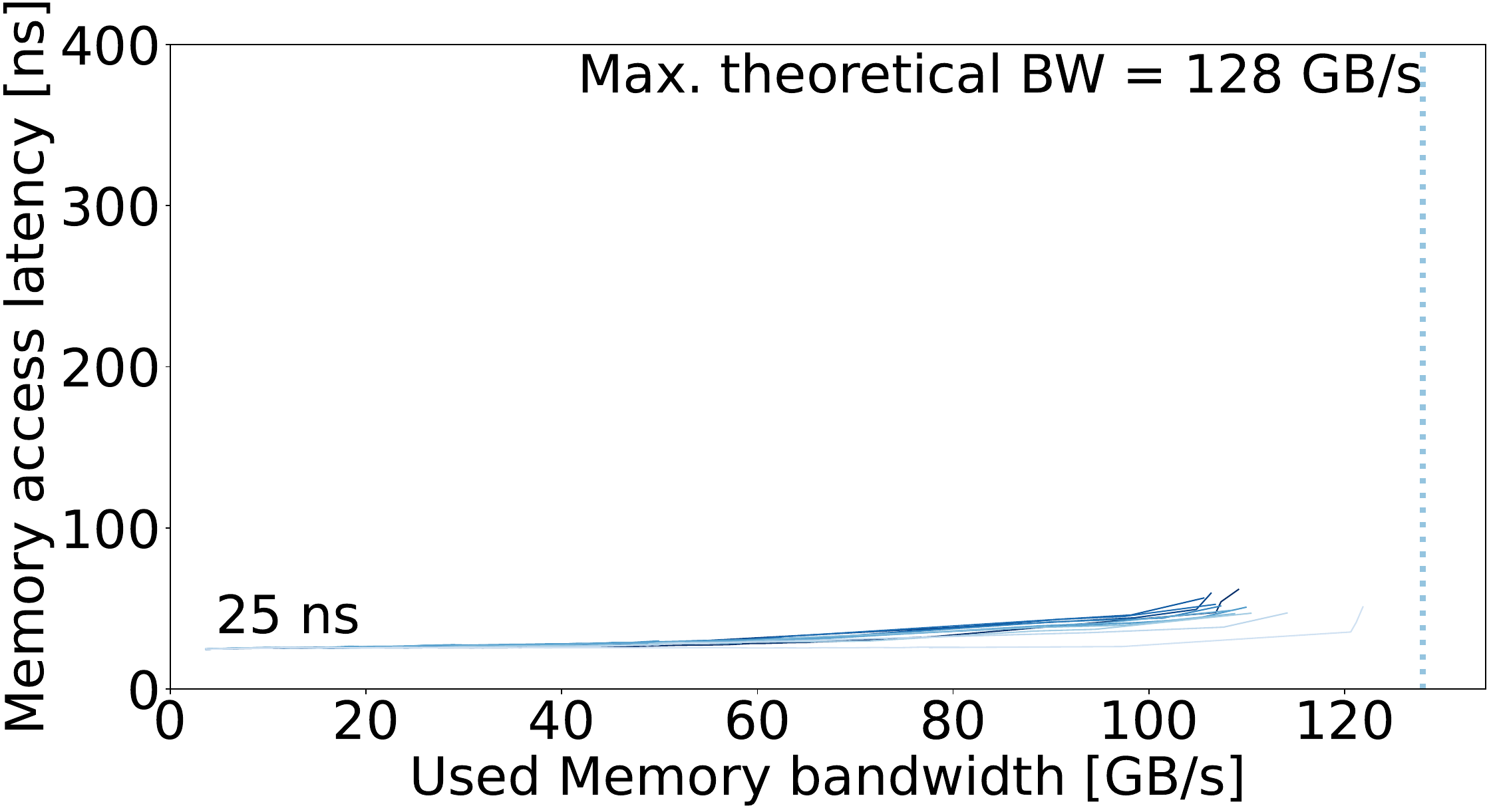}%
  }%

  \caption[Four subfigures]{%
    The updated interface removes the DAMOV \texttt{freqRatio} rounding error. 
    The maximum simulated bandwidth now matches the bandwidth reported by Ramulator (Figure\,\ref{fig:baseNoFix_ramulator}). The memory latency in the application view is still decoupled from the detailed memory simulation: it starts at 25\,ns and increases only mildly as the memory system approaches saturation.}
  \label{fig:baseline-clockfixed}
\end{figure}

\looseness -1 Enabling the clock-scaling code corrected the excessive memory-system bandwidth, 
see Figure\,\ref{fig:baseline-slightly-modify}. This correction, however, did not eliminate all memory-bandwidth discrepancies: in the memory-interface view\,(Figure\,\ref{fig:baseline-slightly-modify-interface}) all experiments show approximately 25\% lower bandwidth than the memory-simulator view (Figure\,\ref{fig:baseNoFix_ramulator}).   

We found that this second cross-simulator clocking issue is caused by the clocking mechanism implemented in the DAMOV interface.   
The interface decides whether to call the memory simulator based on the calculated ratio between the CPU and memory frequencies, \texttt{freqRatio} in Code Listing\,\ref{fig:interface-freq-model}(a).   
The inherent problem with this approach is that \texttt{freqRatio} must be an integer, which DAMOV obtains by rounding up the actual CPU-to-memory frequency ratio.  
Because of this rounding error, the memory frequency is reduced by 2$\times$ instead of 1.57$\times$, leading to a proportional memory-bandwidth discrepancy.


To address this issue, we develop a cross-simulator clocking approach, shown in 
Code Listing~\ref{fig:interface-freq-model}(b), that explicitly tracks simulated CPU and memory time.
The approach first increments the simulated CPU cycle and time counters (lines~1 and~2).
Then, while the simulated memory time lags behind the CPU time (line~3), the interface calls the memory-simulator's tick function and increments the DRAM time and cycle counters (lines~4--6).
The simulated CPU and memory times are incremented by intervals corresponding to their clock periods: 476\,ps ($1/2.1\,\mathrm{GHz}$) for the CPU and 750\,ps ($1/1.333\,\mathrm{GHz}$) for DRAM.
The updated interface removes the DAMOV rounding error, and the maximum simulated memory bandwidth in Figures~\ref{fig:correct-tick-interface} and~\ref{fig:correct-tick-core} now matches the bandwidth provided by Ramulator (Figure~\ref{fig:baseNoFix_ramulator}).

Restoring time-domain consistency also corrects the \textbf{memory-access latency} observed in the memory-interface view. With cross-simulator clocking disabled (Figure~\ref{fig:original-damov-interface}), 
the ZSim interface interprets Ramulator’s cycle time as 476\,ps, matching the CPU clock, instead of 
750\,ps, matching the DRAM clock. 
The memory-interface view therefore reports an artificially low memory latency, starting at only 28\,ns (Figure~\ref{fig:baseNoFix_interface}). 
Enabling the clock-scaling block removes this underestimation, but exposes a second problem:  
DAMOV rounding error  
makes the effective DRAM cycle longer than intended. As a result, the memory-interface view now reports excessively high memory-access latency, starting at 56\,ns (Figure\,\ref{fig:baseline-slightly-modify-interface}). 
After correcting this rounding error, the memory-interface latency closely tracks the memory-simulator view and is increased, as expected, by the queuing delay between the ZSim memory interface and Ramulator’s memory-controller implementation (Figure~\ref{fig:correct-tick-interface}).

The correction of time-domain consistency between ZSim and Ramulator is not a major contribution of this work. Once identified, the underlying issues are relatively straightforward to fix. We nevertheless present this correction first because it places the actual-system measurements and the different simulated memory-performance views within the same, correct bandwidth range. This makes it easier to evaluate the impact of the ZSim--Ramulator memory-model mismatch discussed in 
Section~\ref{sec:CPU–Memory-Model-Mismatch} and the additional simulation enhancements presented in Section~\ref{sec:Further-enhance}.


\begin{takeawaybox}
\textbf{Finding~2.}
Cross-simulator clocking provides an important cautionary example: a seemingly minor integration issue, present in the repositories of the original DAMOV framework~\cite{Oliveira:damov-ZSim-ramulator} and studies based on it~\cite{esmaili:mess, bostanci:ramulator-damov-safari}, can lead to substantial performance inaccuracies. 
Moreover, such an issue may remain undetected if validation relies only on memory-simulator 
statistics~\cite{bostanci:ramulator-damov-safari}. 
\end{takeawaybox}

\newcounter{tempFigure}
\setcounter{tempFigure}{\value{figure}}
\setcounter{figure}{0}

\begin{lrbox}{\codebaseline}
\begin{minipage}{0.254\textwidth}
\begin{lstlisting}[basicstyle=\ttfamily\sixpt, frame=single,framerule=0.8pt, rulecolor={\color[RGB]{220, 20, 60}},numbers=left, numberstyle=\tiny, language=C++]
freqRatio = ceil(cpuFreq/memFreq);
if((tickCounter % freqRatio) == 0)
    wrapper->tick();
tickCounter++;         
\end{lstlisting}
\end{minipage}
\end{lrbox}

\begin{lrbox}{\codeupdated}
\begin{minipage}{0.201\textwidth}
\begin{lstlisting}[basicstyle=\ttfamily\sixpt, frame=single,framerule=0.8pt, rulecolor={\color[rgb]{0.0, 0.4, 0.0}}, numbers=left, numberstyle=\tiny, language=C++]
curCycle++;           
cpuPs += cpuPsPerClk;           
while (cpuPs > dramPs) {      
    wrapper->tick();       
    dramPs += dramPsPerClk;
    dramCycle++;
}
\end{lstlisting}
\end{minipage}
\end{lrbox}

\begin{figure}[!t]

    \renewcommand{\figurename}{Listing} 
    \renewcommand{\theHfigure}{listing.\arabic{figure}}
    \centering
    \subfloat[DAMOV clocking mechanism: \texttt{freqRatio} must be an integer, which DAMOV obtains by rounding up the actual CPU-to-memory frequency ratio.     
    \label{lst_baseline}]{\raisebox{0.38cm}{\usebox{\codebaseline}}}
    \hfill
    \subfloat[Updated interface \label{lst_updated}]{\usebox{\codeupdated}}
    \caption{The updated interface eliminates the DAMOV \texttt{freqRatio} rounding error.}
    \label{fig:interface-freq-model}
\end{figure}

\setcounter{figure}{\value{tempFigure}}


\subsection{Problem\,2: Memory Model Mismatch} 
\label{sec:CPU–Memory-Model-Mismatch}

\subsubsection{\textbf{Initial assessment}}  

The most surprising result in Figures~\ref{fig:original-damov-interface},~\ref{fig:baseline-slightly-modify} and~\ref{fig:baseline-clockfixed} concerns the main-memory latencies observed in the application view. These latencies are expected to exceed those in the memory-interface view by \textbf{tens of nanoseconds}, corresponding to the time memory requests spend in the simulated core, cache hierarchy, and network-on-chip.
However, they start at only 25\,ns and increase only mildly as the memory system approaches saturation.\footnote{In Figures~\ref{fig:baseNoFix_core}~and~\ref{fig:correct-tick-core}, the memory latency starts at 25\,ns. In Figure~\ref{fig:baseline-slightly-modify-core}, it starts at 27\,ns. This small increment is caused by the somewhat higher latency in the memory-interface view (56\,ns) relative to the other experiments. The impact of the memory-simulator and memory-interface latencies on the application view is detailed in Section~\ref{sec:Detailed-analysis}.}
Across the entire bandwidth range, the memory latencies in the application view remain significantly below those reported by the memory simulator and measured at the CPU--memory interface.

We found only two studies that helped us interpret these findings. 
In 2019, the DRAMsim3 developers published a paper summarizing their experience with the ZSim+DRAMsim3 integration~\cite{Shang:misscycle}. 
One of the topics discussed in that paper is the inherent inflexibility of cycle-accurate memory models and the challenges of integrating them with parallel processor simulators. 
The paper also indicates that ZSim integrated with DRAMsim3 may incorrectly relax actual dependencies between consecutive memory requests, effectively reducing simulated memory-access latency. 
The study concludes that the ZSim and DRAMsim3 memory models are incompatible, preventing a correct integration of the two simulators. 
In 2022, an Intel paper co-authored by contributors to the Sniper simulator corroborated this message: loosely synchronized parallel simulators, such as Sniper, do not support cycle-accurate memory simulators because of their delayed response, but instead require an immediate estimate of memory latency~\cite{Eyerman:IRMM}. 

\begin{takeawaybox}
\textbf{Finding~3.}
Previous studies identified a fundamental incompatibility between loosely synchronized 
parallel CPU simulators and delayed-response cycle-accurate memory models.
\end{takeawaybox}

\subsubsection{\textbf{Detailed analysis}} 
\label{sec:Detailed-analysis}

Our study provides a detailed analysis of the ZSim+Ramulator memory-model incompatibility. 
Unlike previous studies that focus on immediate-response memory models~\cite{Eyerman:IRMM, Shang:misscycle}, we show that the incompatibility with delayed-response memory simulators can be addressed through enhancements to the CPU--memory simulator interface. 
This result enables the integration of cycle-accurate memory simulators into massively parallel simulation infrastructures.

Understanding the ZSim+Ramulator memory-model mismatch requires some knowledge of the overall ZSim simulation flow. 
To speed up simulation, ZSim partitions application execution into 1000-cycle windows and processes each window in two phases. 
In the first phase, the Bound phase, each core is simulated independently using embarrassingly parallel multi-core simulation. 
This phase models in detail the instructions that have no dependencies on other cores, such as arithmetic operations and cache accesses.

The Bound phase cannot model memory timings because they depend on collisions among concurrent memory requests from all cores, information that is not available during this simulation phase.    
This is an important limitation because, in a tightly coupled simulation, instructions that depend on memory requests can be issued only after those requests are resolved. 
This would introduce dependencies between simulated cores, and between core and memory simulation, effectively defeating the purpose of embarrassingly parallel multi-core simulation. 
ZSim's solution is to relax these dependencies and issue the dependent instructions after a predefined delay.  
These issue timings are only tentative and are corrected in the following simulation phase.

Indeed, the second simulation phase, the Weave phase, models in detail the latencies of memory requests issued by all cores and uses these latencies to correct the tentative issue timings of dependent instructions. 
The overall approach works well and provides fast simulation of multi-core systems, making ZSim one of the most widely used simulators for computer-architecture exploration~\cite{Akram:simulationSurvey,Akram:haswel,zsim:repo}.
The Weave-phase adjustments of issue timings work correctly for instructions executed in the CPU core and caches. 
However, they do not properly adjust dependent memory accesses, which is a major limitation of this simulation platform.  
%
Although the integrated ZSim+Ramulator simulation platform is publicly available and actively used in computer-architecture research~\cite{Yousefijamarani:aiAccelerator,Attia:damovCryptoAccel,kazi:damov,olivera:thesisdamov,DL-PIM:damov,shokouhinia:damov,Ghiasi:ALPDamov}, its memory-model incompatibility is not well known or understood by the community. For this reason, we explain it in detail. 
 
\begin{figure}[!t]
  \centering
    \includegraphics[width=1\linewidth]{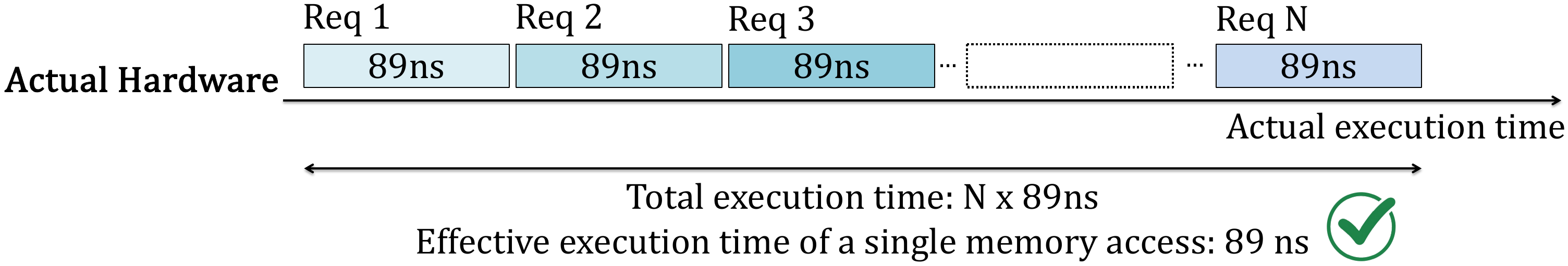}%
  \caption[Four subfigures]{%
  In actual systems, dependent memory requests are issued only after the previous request is resolved.} 
  \label{fig:ptr-chase-timeline-hw}
\end{figure}

\begin{figure}[!t]
  \centering
    \hspace{-.4cm}\includegraphics[width=1.03\linewidth]{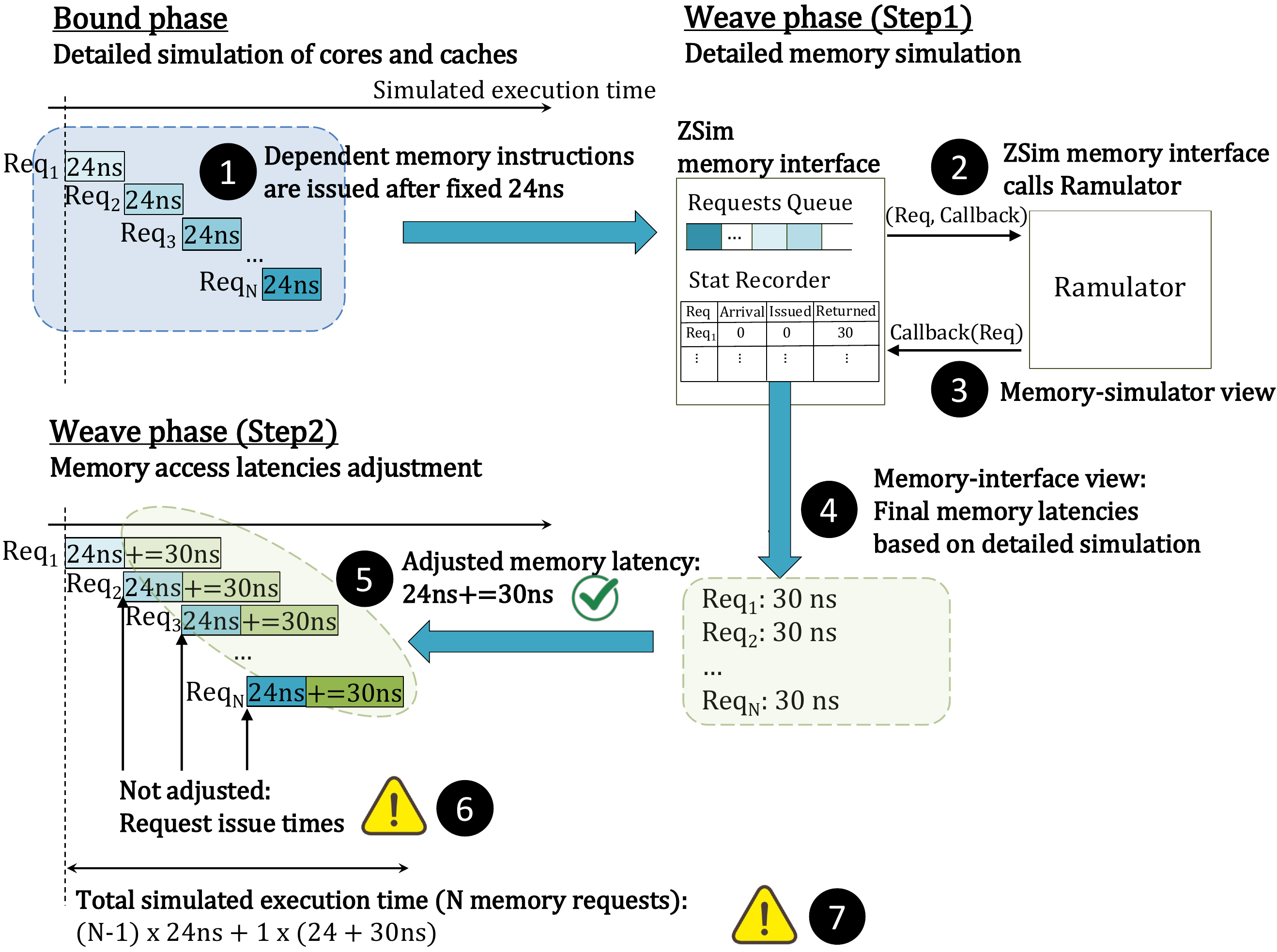}%
    
  \caption[Four subfigures]{%
  In the ZSim Bound phase, instructions dependent on memory operations are issued after 24\,ns\,\circled{1}. These issue times are adjusted in the Weave phase based on detailed memory simulation\,\circled{2}\circled{3}\circled{4}\circled{5}. 
  The problem is that dependent memory instructions have already been passed to Ramulator, so their issue timings cannot be adjusted\,\circled{6}. This has a severe impact on their simulated execution time\,\circled{7}.} 
  \label{fig:ptr-chase-timeline-baseline}
\end{figure}

Figures~\ref{fig:ptr-chase-timeline-hw} and~\ref{fig:ptr-chase-timeline-baseline} compare actual execution with the ZSim simulation of a sequence of dependent memory accesses. 
In the actual hardware, each request can be issued only after the previous one is resolved, 
and the total execution time is the sum of the individual memory latencies. 

%


Figure~\ref{fig:ptr-chase-timeline-baseline} shows the simulation of the same sequence of memory accesses in the ZSim+Ramulator infrastructure.
The first ZSim simulation phase (Bound phase) uses an immediate-response fixed-latency memory model. In the DAMOV interface, this latency is set to a single CPU cycle~\cite{safari:damov-template}.
This means that an instruction dependent on a memory access is issued after a delay comprising the time the memory operation spends in the simulated core, cache hierarchy and network-on-chip, plus the one-cycle memory-model latency.
ZSim determines this issue delay based on the simulated architecture. For our baseline architecture (Table~\ref{table:configuration}), the resulting inter-issue delay is 24\,ns\,\circled{1}.

In the second phase, the Weave phase, the ZSim memory interface calls Ramulator to perform detailed memory simulation\,\circled{2}. 
The memory interface then receives the results from Ramulator, corresponding to the memory-simulator view\,\circled{3}, and adds the queuing delay between the CPU memory interface and Ramulator's memory-controller implementation\,\circled{4}. 
These latencies, obtained from detailed memory simulation, are used to adjust the tentative latencies from the Bound phase: 
\mbox{$Req_i~latency = 24\,ns + 30\,ns$}\,\circled{5}.  
Detailed memory simulation, however, \textbf{does not correct the issue times of dependent memory instructions}, $Req_2,~Req_3,\ldots$\,\circled{6}, because these instructions are passed to Ramulator before the preceding 
memory-request timings are adjusted.\footnote{We present a simplified example that illustrates the memory-model incompatibility between ZSim and Ramulator. 
The full complexity of the synchronization between Bound and Weave simulation phases exceeds the scope of this study. }  
This has a profound implication for how the simulated application perceives main-memory latency\,\circled{7}.

\begin{takeawaybox}
\textbf{Finding~4.}
The simulated execution time of any sequence of dependent memory operations, e.g., in the pointer-chase benchmark, is determined primarily by the \textbf{inter-issue time} between these instructions (24\,ns), while the simulated memory-access latency affects execution time only through the last memory access in the sequence. 
For this reason, memory-access latencies simulated by Ramulator have only a minor impact on the memory latency perceived by the simulated application.
\end{takeawaybox}

\subsubsection{\textbf{Solution}}
\label{sec:model-matched}

\begin{figure}[!t]
  \centering
    \hspace{-.1cm}
    \includegraphics[width=1.01\linewidth]{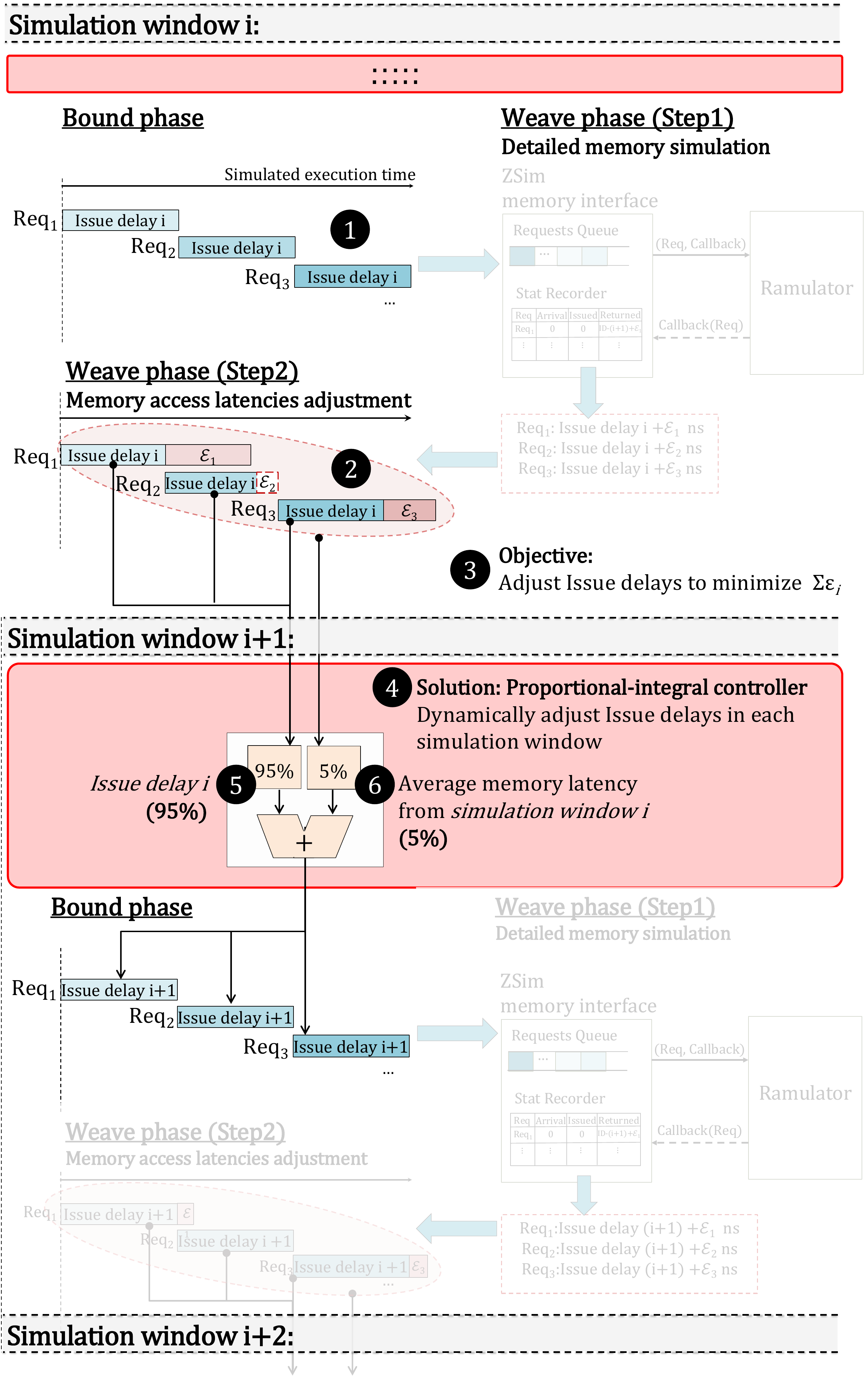}%
    \vspace{-.1cm}
  \caption[Four subfigures]{%
The smaller the difference $\varepsilon_i$ between the issue delay\,\circled{1} and the simulated memory latency\,\circled{2}, the tighter the coupling between the application view and the detailed memory simulation. 
To minimize the overall simulation error $\Sigma\varepsilon_i$\,~\circled{3}, in each simulation window, a proportional--integral controller dynamically adjusts the issue delay\,\circled{4}\circled{5}\circled{6}.
} 
  \label{fig:ptr-chase-timeline-proposal}
\end{figure}

The essence of the problem is the premature issue of dependent memory instructions after a fixed 24\,ns delay. 
Figure~\ref{fig:ptr-chase-timeline-proposal} illustrates our solution using the example of simulation windows~\emph{i},~\emph{i+1}, and~\emph{i+2}. 
The process is repeated in each simulation window and for each simulated core.
The approach is based on the observation that the smaller the difference $\varepsilon_i$ between the issue delay in the Bound phase\,\circled{1} and the simulated memory latency in the Weave phase\,\circled{2}, the smaller the gap between the application memory-performance view and the detailed memory simulation. 
Therefore, our objective is to minimize the cumulative difference $\Sigma\varepsilon_i$ over the whole simulation\,\circled{3}.

\looseness -1 We address this problem with a proportional--integral controller mechanism from classical control theory~\cite{Franklin:PIDController}, which dynamically adjusts the issue delay at the beginning of each simulation window\,\circled{4}. 
The controller uses a feedback mechanism that combines the previous issue-delay estimate (95\% weight)\,\circled{5} and the average memory latency from the previous cycle-accurate memory-simulation phase (5\% weight)\,\circled{6}.
The larger the difference between the issue delay and cycle-accurate memory latency, and the higher the user-defined correction factor (5\% in our implementation), the larger the controller correction. 
The objective is not to reach the correct issue delay in a single iteration, but to progressively reduce accumulated error and remove persistent bias over time~\cite{Franklin:PIDController,Goodwin:ControlSystemDesign}.

\begin{figure}[!t]
  \centering
  \includegraphics[width=.9\linewidth]{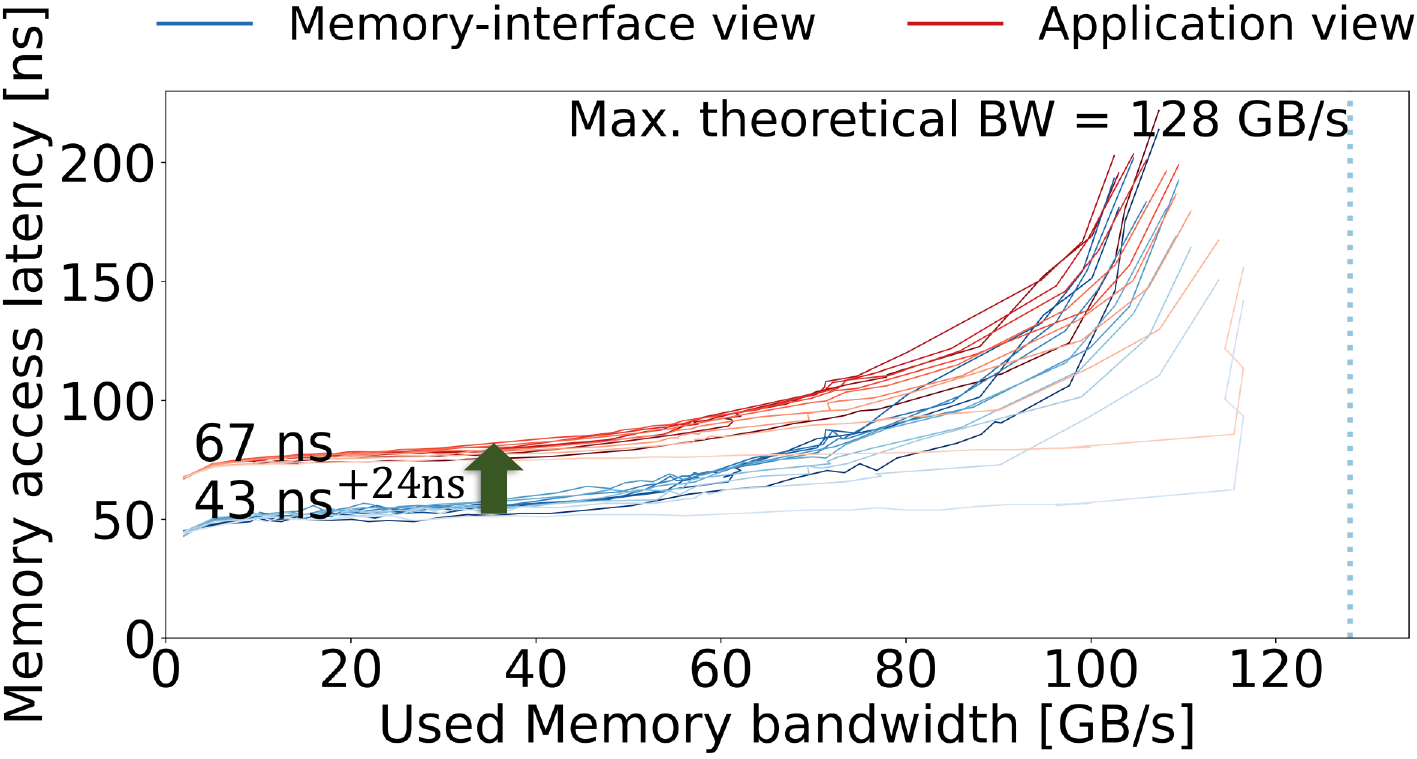}%
  \caption[Four subfigures]{%
  After correcting the memory-model mismatch, the application view closely follows the interface view. 
  The 24\,ns latency increment corresponds to the time memory requests spend in the simulated core, cache hierarchy, and network-on-chip.} 
  \label{fig:correct-interface-virgin}
\end{figure}

Figure~\ref{fig:correct-interface-virgin} shows the application view of memory performance after this correction (red curves). 
As a reference, the figure also plots the memory-interface view (blue curves). 

\begin{takeawaybox}
\textbf{Finding~5.}
After correcting the memory-model mismatch, the application view closely follows the interface view across the full range of simulated memory bandwidths. 
The 24\,ns latency increment corresponds to the time memory requests spend in the simulated core, cache hierarchy, and network-on-chip. 
The memory latency perceived by simulated applications is now \textbf{coupled to the detailed memory simulation}. 
\end{takeawaybox}

\section{Further enhancements}
\label{sec:Further-enhance}

Fixing the ZSim+Ramulator memory-model incompatibility and coupling the application memory view to the detailed memory simulation, however, does not resolve all discrepancies between simulated and actual memory-system performance. 
The simulated latencies in Figure~\ref{fig:correct-interface-virgin} remain significantly below the actual measurements. 
In this section, we briefly present the simulator enhancements we deploy to make simulated memory performance resemble actual-system behavior.


\subsection{Address mapping} 
\label{sec:Address-mapping}

\looseness -1 The next correction addresses the performance impact of read and write memory traffic. 
In the actual system, 100\%-read traffic achieves the lowest latency and highest bandwidth. 
Memory writes reduce performance and cause the memory system to reach saturation sooner. 
The higher the percentage of writes, the larger the performance impact; consequently, the actual system shows a clear gradient from the lightest curve (100\%-read traffic) to darker memory curves, as shown in Figure\,\ref{fig:actual-hw}. 
This trend is not fully reflected in Figure\,\ref{fig:correct-interface-virgin}.\footnote{In the initial simulation infrastructure, the memory simulator showed the same trend as the actual hardware, but this behavior was affected by the changes to the time-domain consistency and the memory model.} 
We found that this discrepancy originates from the memory simulator's simplified \textbf{mapping between physical addresses and actual data locations}: memory channel, DIMM, rank, bank, row, and column. 
After deploying a complex mapping reverse-engineered from actual systems~\cite{Wang:addressMapping}, memory curves with different read/write traffic compositions closely follow the actual trends, as shown in Figure\,\ref{fig:final-am-core}. 
Our study is the first to show that simplified address mapping can cause an incorrect ordering of bandwidth–latency curves with different read/write ratios.   
The latest versions of memory simulators~\cite{Haocong:Ramulator2,steiner:dramsys4} already support more advanced address mappings, and we strongly recommend that users configure them in their simulations. 
%

\subsection{Network-on-chip}  

Load-to-use latency is also affected by the CPU \textbf{network-on-chip~(NOC)}.   
Our baseline simulation platform uses a simple NOC model with a fixed delay included in the overall last-level cache latency~\cite{esmaili:acm}.
We replace the initial ZSim NOC with a more advanced M/D/1 model~\cite{Oliveira:damov-ZSim-ramulator} and configure it to closely match the Intel Xeon Platinum 8160 processor~\cite{intel_xeon_8160}.    
The configuration is based on detailed analyses of the processor architecture~\cite{dai:reverseNOC} and latencies~\cite{sodani:noclatency}, as well as the mapping between cores and their corresponding network tiles~\cite{mccalpin:mapping}.  
This adds 10\,ns to memory latency across the full bandwidth range, as shown in Figure\,\ref{fig:final-addnoc-core}.

\subsection{Prefetching}  
Finally, \textbf{data prefetchers} have an important impact on memory-traffic intensity.  
We implement stride prefetchers in ZSim cores. 
This increases overall memory traffic, resulting in up to 37\,ns higher saturated memory latency (Figure\,\ref{fig:prefetcher}).

\begin{takeawaybox}
\textbf{Finding~6.}
Achieving close-to-hardware memory-simulation
accuracy also requires correct address mappings, detailed
network-on-chip models and data prefetchers.
\end{takeawaybox}


\section{\mbox{Final evaluation and further improvements}}
\label{sec:Quantitative-validation}


\subsection{\mbox{ZSim+Ramulator:\,Pointer-chase\,and\,STREAM}}  

Sections~\ref{sec:Init-validation},~\ref{sec:Integration}, and~\ref{sec:Further-enhance} present several simulator enhancements and evaluate them using detailed bandwidth--latency curves. 
In this section, we complement the analysis with a quantitative evaluation using standard \textbf{pointer-chase}~\cite{Verdejo:microbenchmarks,sanchez:zsimEnhance} and \textbf{STREAM}~\cite{mccalpin:streamLink, mccalpin:streamBenchmark} benchmarks.   
Figure~\ref{fig:memory-intensive-ramulator} shows the relative differences between actual and simulated performance. 
Overall, as expected, the pointer-chase and STREAM results are consistent with the application view 
in Figures\,\ref{fig:original-damov-interface},\,\ref{fig:baseline-slightly-modify} and\,\ref{fig:baseline-clockfixed}  
because this view ultimately determines simulated application performance. 

The \textbf{initial simulation infrastructure} shows unrealistically high STREAM bandwidths and low pointer-chase latencies relative to actual memory-system performance. 
The resulting simulation errors are large, ranging from -72\% to 77\%.

Correcting \textbf{time-domain inconsistencies} led to significant improvements in simulated memory bandwidth. 
After these corrections, the simulated STREAM kernel results closely match the actual measurements, with an average error of 8\%.  
The difference between simulated and actual pointer-chase latency remains high, at -72\%.
 
Correcting the ZSim+Ramulator \textbf{memory-model inconsistencies} 
further improved the accuracy of STREAM simulations, reducing the average error to less than 4\%.  
More importantly, this correction effectively couples the memory latency perceived by simulated applications to the detailed memory simulation. This reduces the pointer-chase error from -72\% to -28\%.

\begin{figure}[!t]
  \centering
	\includegraphics[width=.8\linewidth]{graphics/horizontal-legend}
\vspace{-.3cm}

  \subfloat[Correct address mapping%
           \label{fig:final-am-core}]{%
    \includegraphics[width=.5\linewidth]{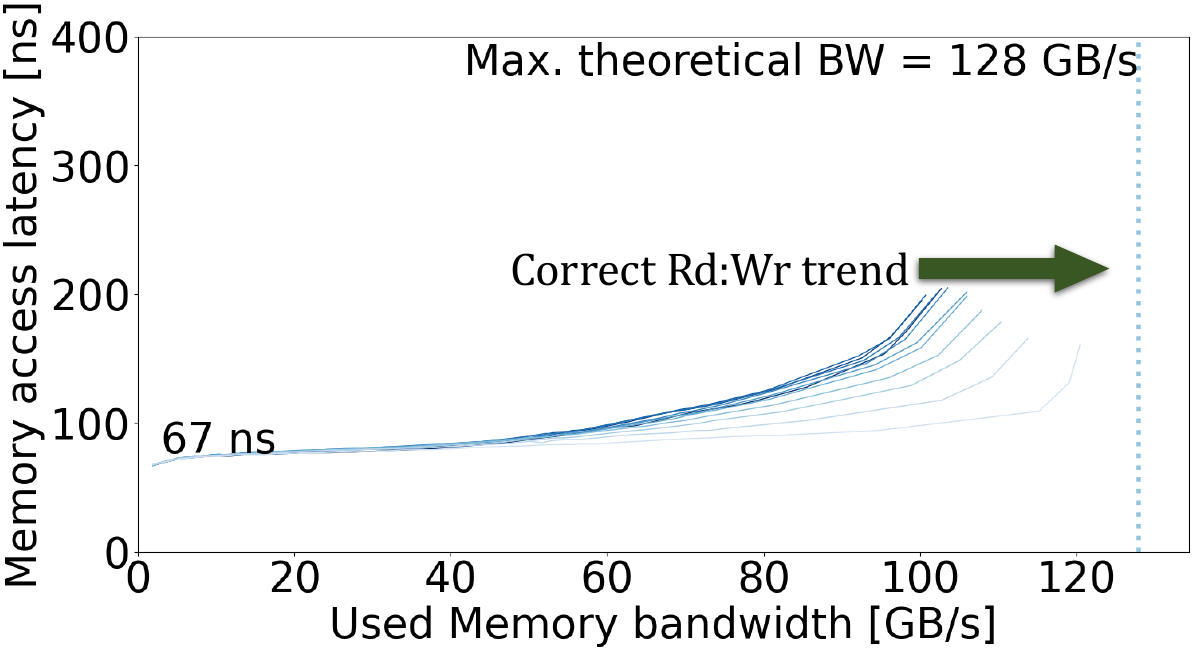}%
  }%
  \subfloat[Realistic network-on-chip%
           \label{fig:final-addnoc-core}]{%
    \includegraphics[width=.5\linewidth]{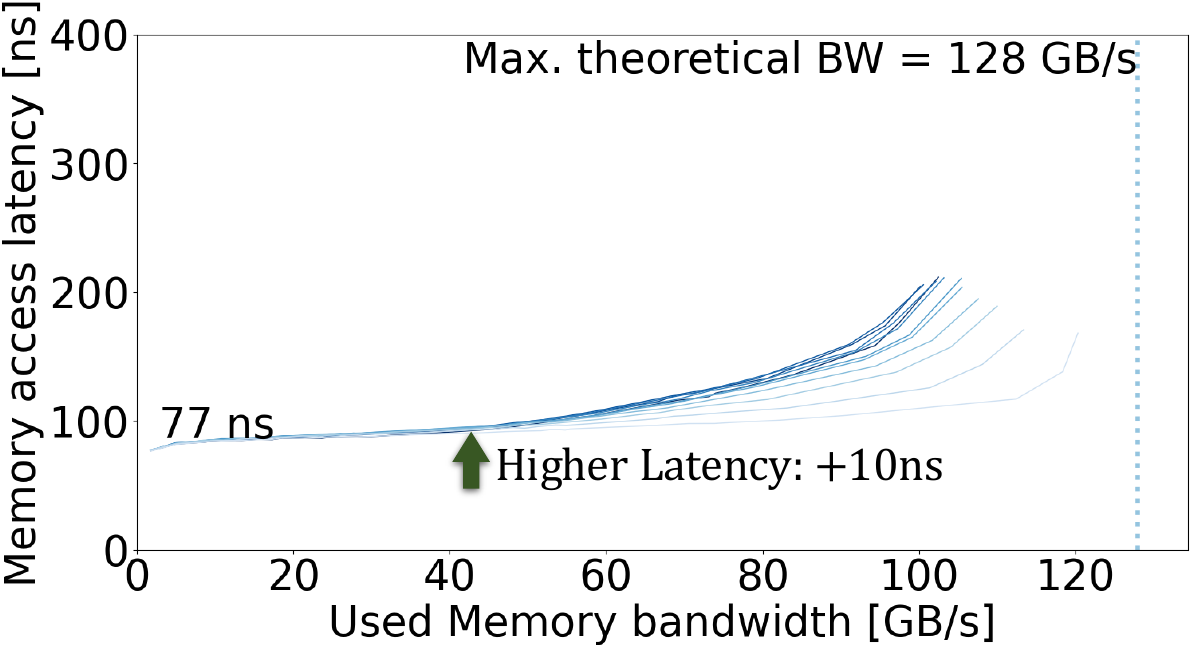}%
  }%
    \vspace{-.2cm}
    
  \subfloat[Adding data prefetchers%
           \label{fig:prefetcher}]{%
    \includegraphics[width=.5\linewidth]{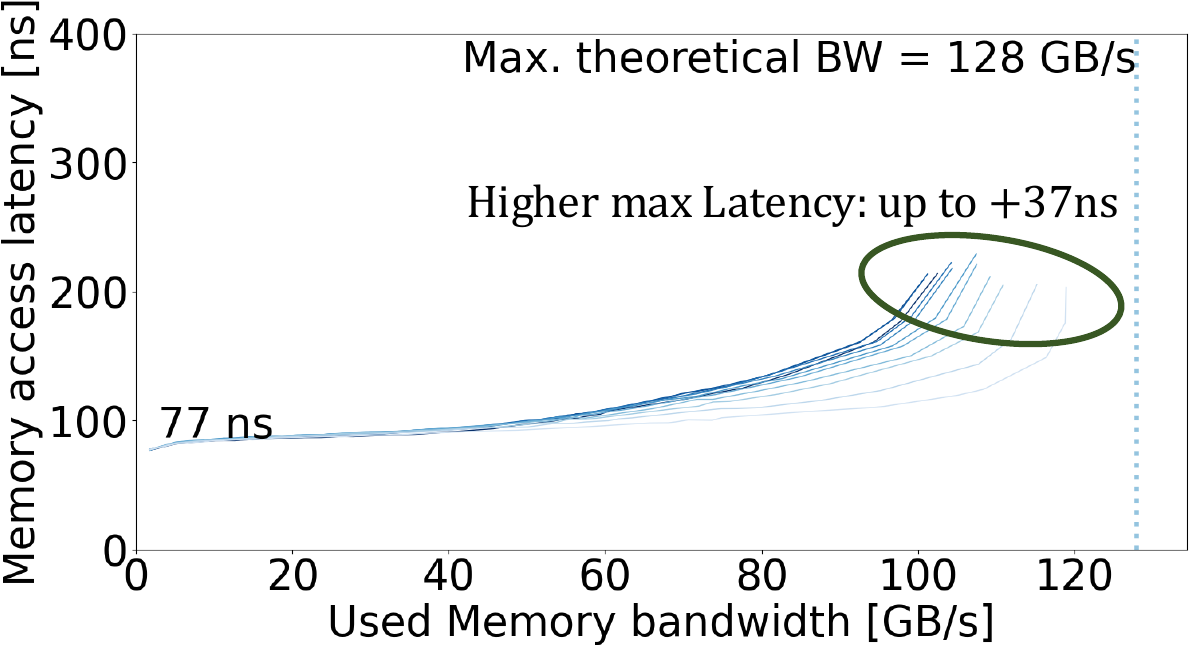}%
  }%
  
  \caption[Three subfigures]{%
  ZSim+Ramulator --- Application view: 
  Achieving close-to-hardware memory-simulation accuracy also requires correct address mappings, detailed network-on-chip models and data prefetchers.}
  \vspace{-.2cm}
  \label{fig:ZSim-ramulator-final-modifications}
\end{figure}

\begin{figure*}[!t]
  \centering
    \includegraphics[width=1\linewidth]{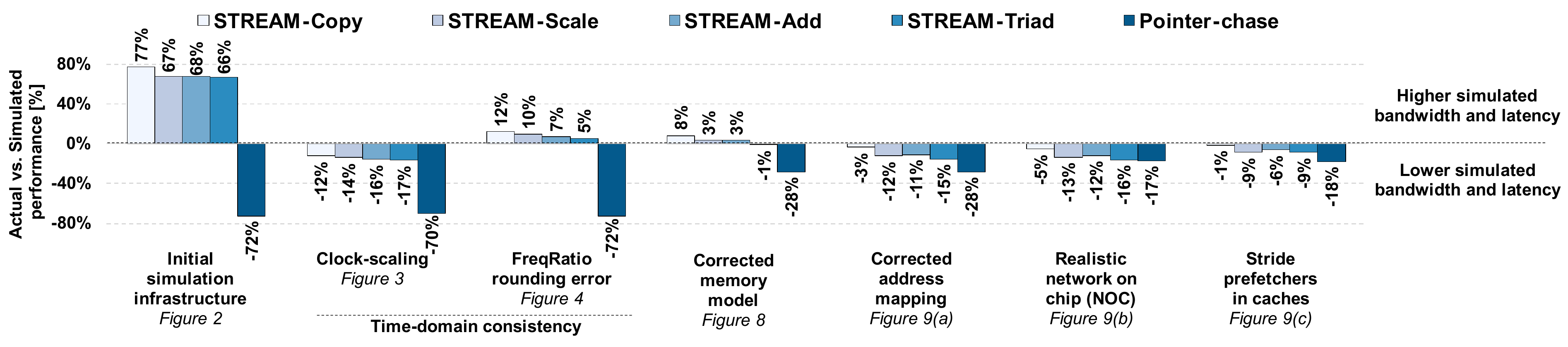}%
   \vspace{-.3cm}
  \caption[Four subfigures]{%
  The pointer-chase and STREAM results are strongly correlated with the application views in the corresponding Figures\,\ref{fig:original-damov-interface},\,\ref{fig:baseline-slightly-modify},\,\ref{fig:baseline-clockfixed},\,\ref{fig:correct-interface-virgin}\,and\,\ref{fig:ZSim-ramulator-final-modifications}.    
  The simulator provides accurate memory-system performance with somewhat lower latency than actual hardware.} 
  \vspace{-.3cm}
  \label{fig:memory-intensive-ramulator}
\end{figure*}

The \textbf{address-mapping corrections} had no impact on the pointer-chase results, while their STREAM evaluation led to an interesting finding. 
The advanced address mapping corrected the ordering of curves with different read/write ratios (Figure~\ref{fig:final-am-core}), thereby improving simulator behavior and memory-performance trends.  
However, this also led to somewhat higher STREAM simulation errors (-10\% average error).  
This is because the address-mapping correction reduced the performance of traffic with moderate and high percentages of memory writes, which affected the STREAM kernels.  
%
This example shows that improving simulator behavior and performance trends does not necessarily reduce simulation inaccuracies for a specific benchmark.

The ZSim extension with a \textbf{realistic NOC module} and its correct configuration increased memory latency by approximately 10\,ns. 
This reduced the pointer-chase simulation error from -28\% to -17\%, while having no major impact on the STREAM results. 

The implementation of \textbf{stride prefetchers in ZSim cores} further improved STREAM accuracy by increasing simulated bandwidth. 
The average error of the STREAM kernels was reduced to only -6\%.  
As expected, stride prefetchers had no impact on the pointer-chase results.

\looseness -1 Overall, the simulator provides accurate memory-system performance, although with somewhat lower latency than the actual hardware. 
Some sources of the remaining simulation inaccuracies and potential corrections are discussed in Section~\ref{sec:Opportunities-further-improvements}.

\subsection{Ramulator\,2, DRAMsim3 and DRAMSys\,4}
\label{sec:finalChart}

\begin{figure}[!t]
\centering
\includegraphics[width=.8\linewidth]{graphics/horizontal-legend}
\vspace{-.3cm}

\subfloat[\mbox{Intel\,Skylake\,CPU\,with\,6xDDR4-2666}%
           \label{fig:gem5-bw-lat-result-characterization-hardware}]{%
    \includegraphics[width=.5\linewidth]{graphics/bandwidth--latency-curve-intel-skylake}%
  }%
  \hfill
  \subfloat[ZSim+Ramulator\,2%
           \label{fig:app-view-ramulator2}]{%
    \includegraphics[width=.5\linewidth]{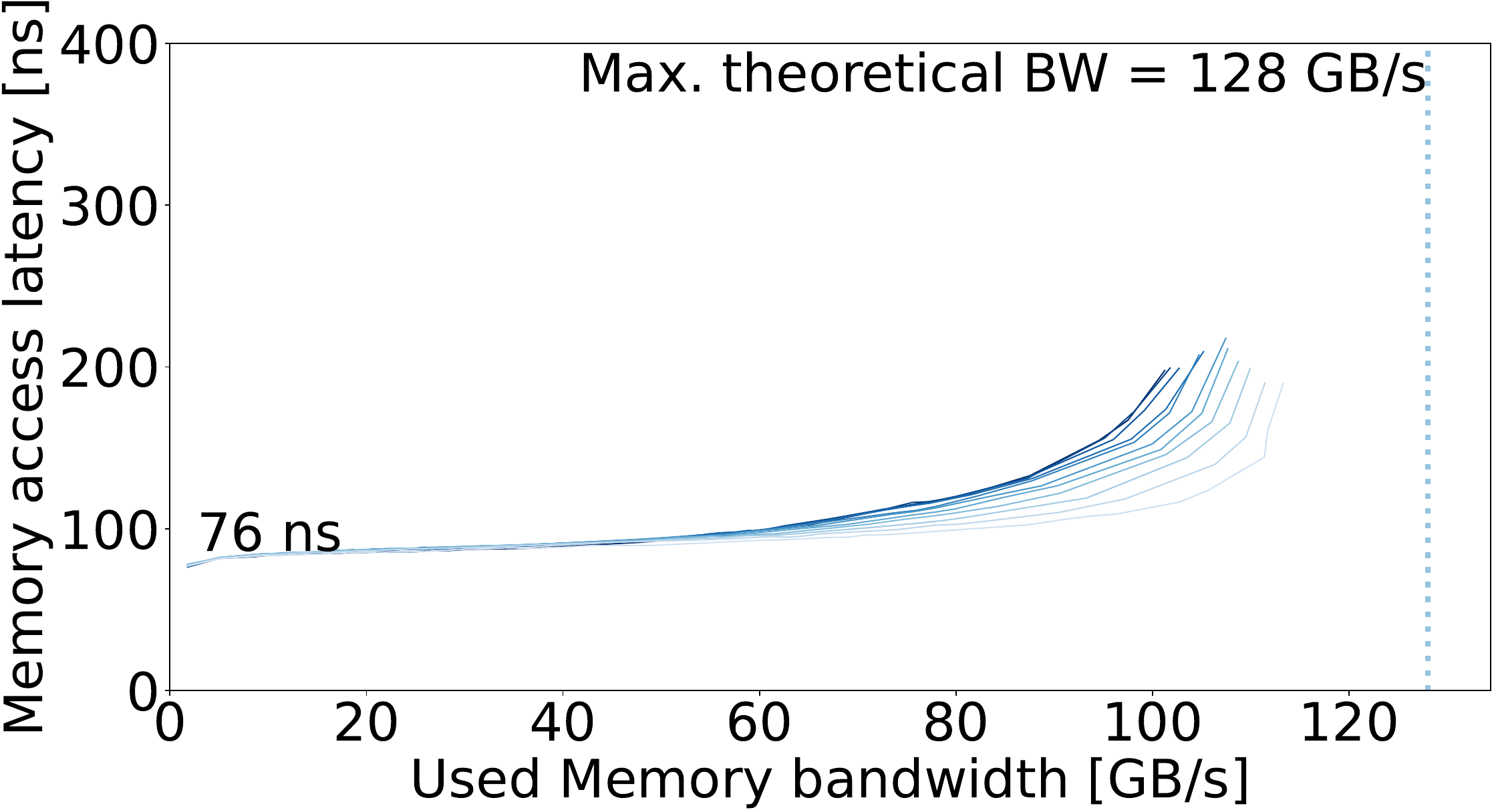}%

  }%

   \vspace{-.2cm}
   
  \subfloat[ZSim+DRAMsim3%
           \label{fig:gem5-bw-lat-result-characterization-ramulator2}]{%
    \includegraphics[width=.5\linewidth]{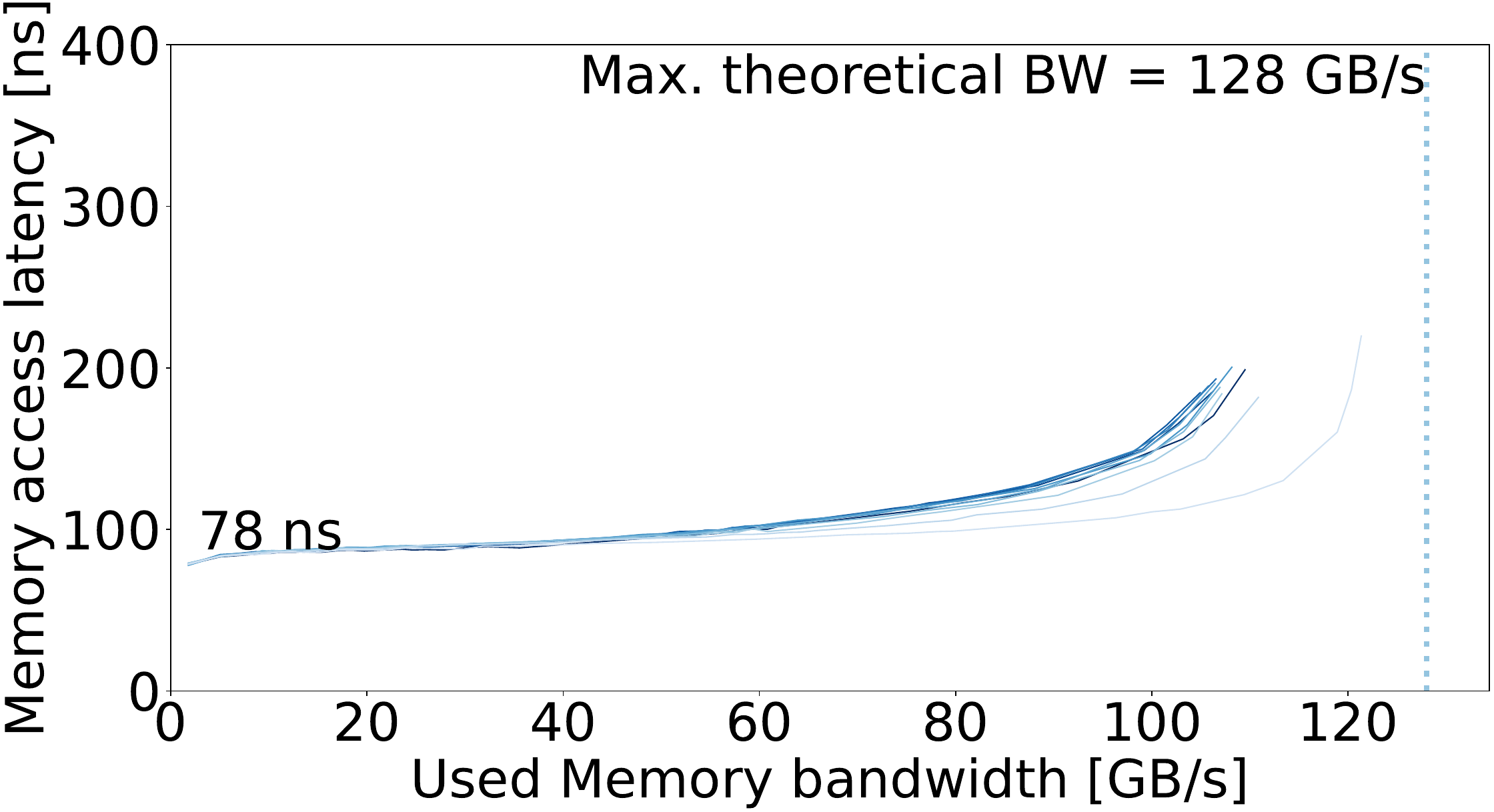}%

  }%
  \hfill
  \subfloat[ZSim+DRAMSys\,4%
           \label{fig:app-view-dramsys}]{%
    \includegraphics[width=.5\linewidth]{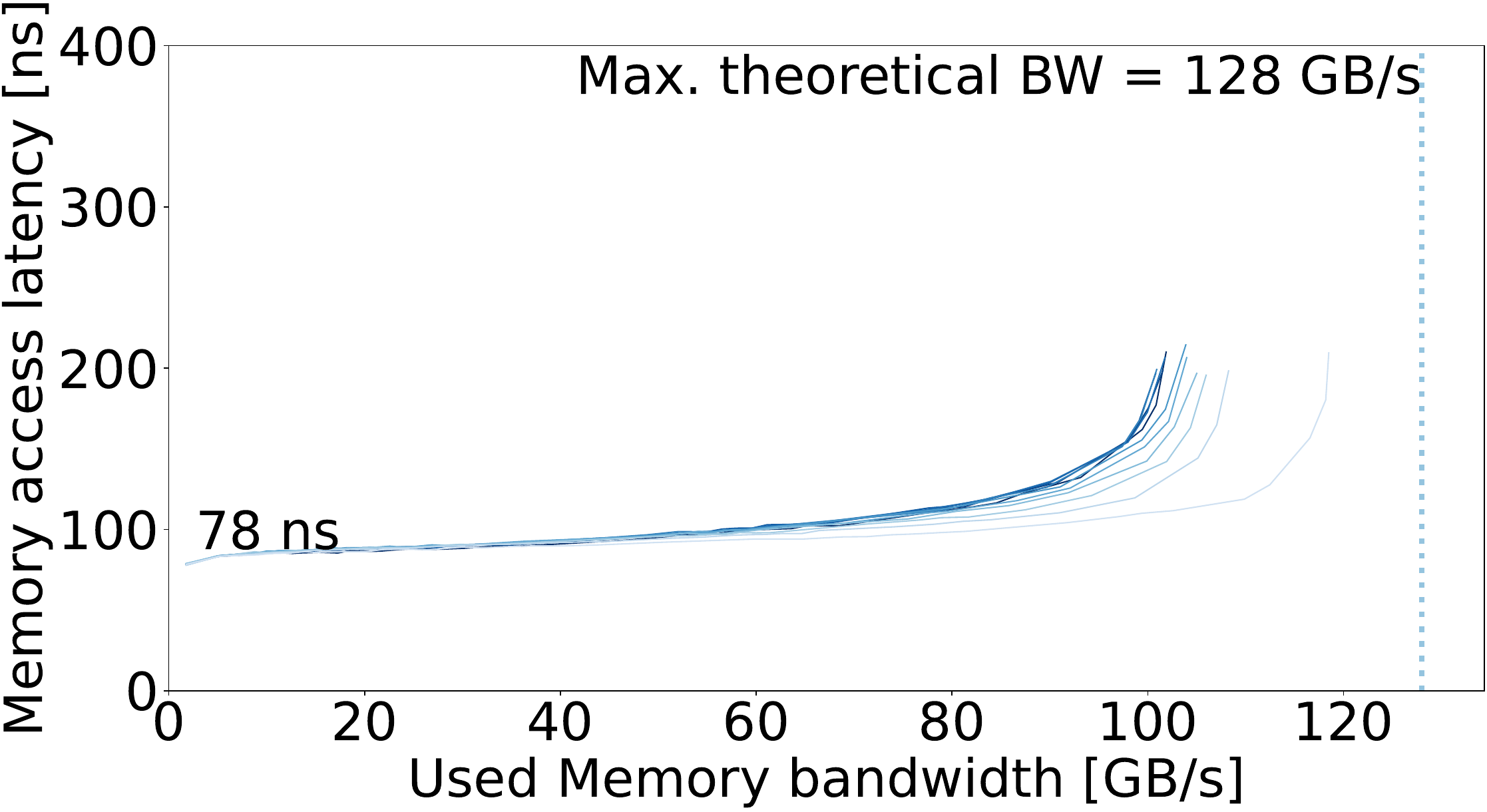}%
  }%

  \subfloat[Pointer-chase and STREAM%
           \label{fig:memory-intensive-all-sim}]{%
    \includegraphics[width=1\linewidth]{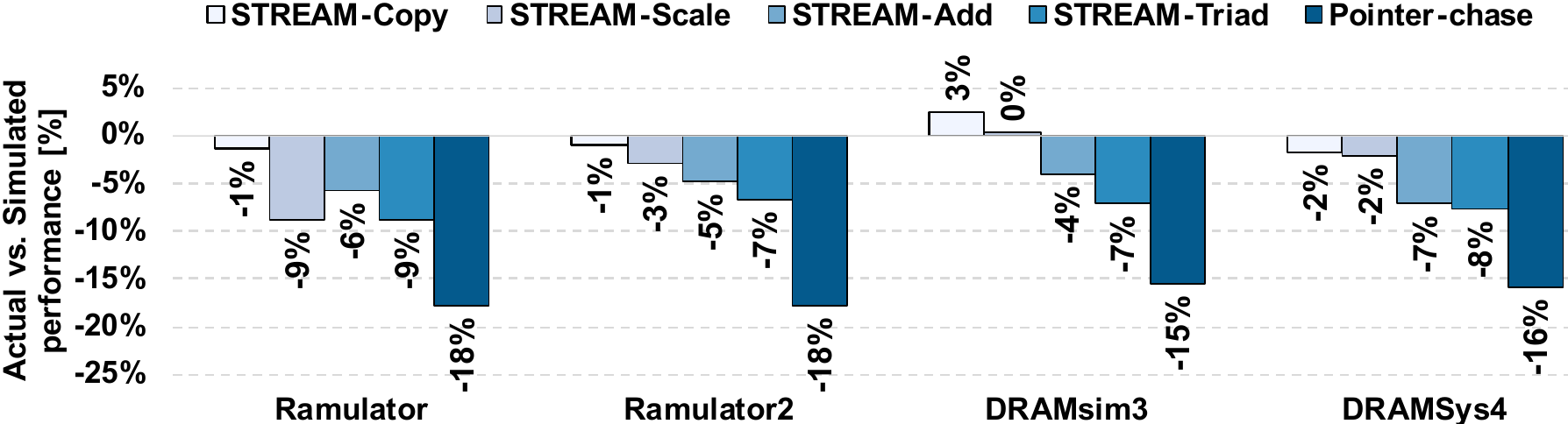}%
  }%

  \caption[Four subfigures]{%
   ZSim connected to Ramulator\,2, DRAMsim3 and DRAMSys\,4 follows the actual-system measurements, and the performance differences among memory simulators are minor. The results also closely match the final ZSim+Ramulator evaluations.} 
   \vspace{-.3cm}
  \label{fig:final-mem-simulators}
\end{figure}

To demonstrate the broader applicability of our enhancements, we deploy them in ZSim connected to Ramulator\,2, DRAMsim3, and DRAMSys\,4, covering all major public cycle-accurate memory simulators. 
Figures~\ref{fig:app-view-ramulator2}--\ref{fig:app-view-dramsys} show detailed bandwidth--latency curves for all simulators, while Figure~\ref{fig:memory-intensive-all-sim} summarizes their simulation errors measured with the pointer-chase and STREAM benchmarks.


\looseness -1 The bandwidth--latency curves of all simulators have very similar shapes and key characteristics. Overall, they also closely follow the shape of the actual-system measurements.
Latency starts at 76--78\,ns and increases only slightly up to approximately 80\,GB/s. Beyond this point, the latency increase becomes more pronounced, and all simulators reach saturation at approximately 100--120\,GB/s.
The maximum simulated latencies under saturation are also similar, ranging from 179\,ns to 227\,ns. In addition, all bandwidth--latency curves show that the higher the proportion of memory writes, the sooner the simulators reach saturation.

Finally, all simulators generally show slightly lower STREAM bandwidths and 15--18\% lower pointer-chase latencies than the actual hardware (Figure~\ref{fig:memory-intensive-all-sim}).

\begin{takeawaybox}
\textbf{Finding~7.}
The Ramulator\,2, DRAMsim3, and \mbox{DRAMSys\,4} results closely match the final ZSim+Ramulator evaluations.  
Notably, the performance differences among the cycle-accurate memory simulators are minor, and substantially smaller than the discrepancies caused by the different simulation setups and corrections discussed in Sections~\ref{sec:Init-validation},~\ref{sec:Integration}~and~\ref{sec:Further-enhance}.
\end{takeawaybox}

\subsection{Opportunities for further improvements}
\label{sec:Opportunities-further-improvements}

\looseness -1  After the corrections, the memory latency of all simulators under study remains below the actual-system measurements.  
In the unloaded memory system, this difference is modest, approximately 13\,ns. 
However, the inaccuracy increases with memory-bandwidth utilization and reaches hundreds of nanoseconds in the saturated region of the bandwidth--latency curves. 

One potential contributor to the remaining latency discrepancy is incomplete modeling of implementation-specific delays in the memory controller and PHY. Although memory simulators model controller behavior, including request arbitration and scheduling, they do not necessarily capture the pipeline delays of a particular controller implementation or the latency introduced by the memory PHY and associated I/O circuitry~\cite{hansson:memctrl}. 
In simulation infrastructures using Ramulator or DRAMsim3, a simple approximation would be to introduce a fixed additional delay through a delay buffer, calibrated so that the simulated application-view unloaded latency matches the corresponding hardware measurement. 
DRAMSys already provides configuration options for forward and backward delays in the memory controller, arbiter and DRAM PHY.  
Such calibration addresses the unloaded-latency discrepancy but does not by itself establish accuracy under load. 
More detailed timing models would require characterization of the target memory-controller pipeline, PHY, and controller-to-DRAM interconnects.



 
Also, 
higher traffic intensities, e.g., those caused by more aggressive data prefetchers, could further saturate the simulated memory systems. 
This could increase simulated STREAM performance, which is still somewhat below the actual performance. 
More importantly, it could substantially increase latencies in the saturated regions of the simulated bandwidth--latency curves, reducing the current discrepancy with the actual measurements.

\begin{figure}[!t]
  \centering
  \includegraphics[width=.9\linewidth]{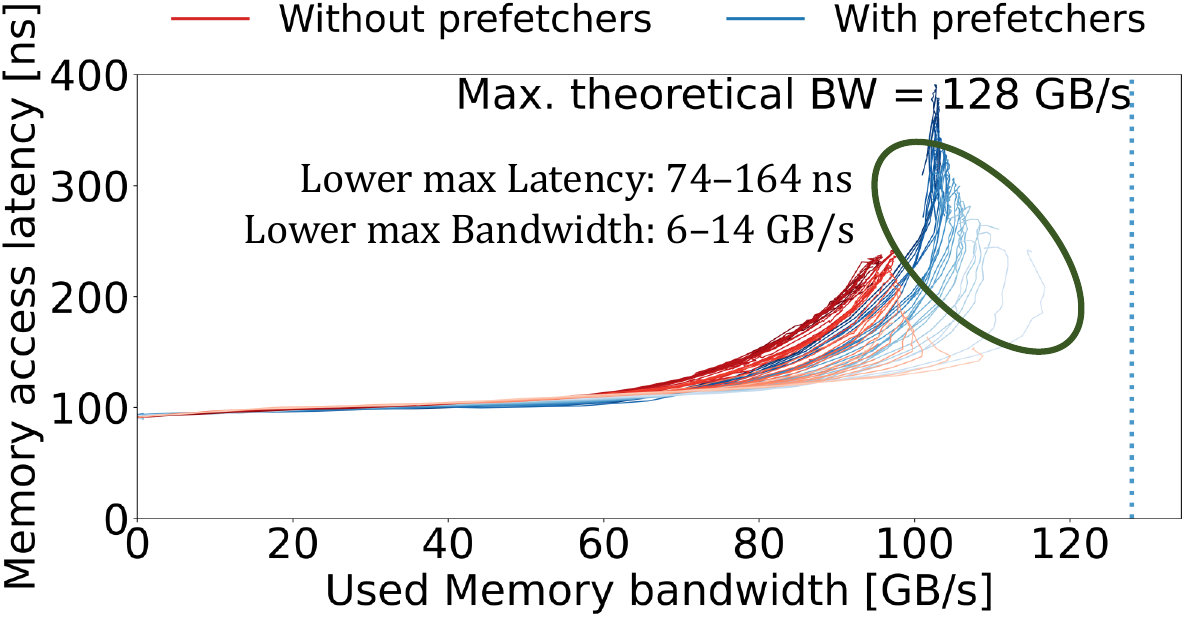}%
  \vspace{-.3cm}
  \caption[Four subfigures]{%
  Data prefetching is important for fully saturating the main memory of the actual Intel Xeon Platinum 8160 system. With prefetchers disabled, the maximum memory bandwidths drop by 6--14\,GB/s, while the maximum latencies decrease by 74--164\,ns.  }
  \vspace{-.4cm}
  \label{fig:hw-prefetcher}
\end{figure}

Figure~\ref{fig:hw-prefetcher} confirms the impact of data prefetching on memory-system saturation in the actual Intel Xeon Platinum 8160 system~\cite{BSC:tech,intel_xeon_8160}.
The figure presents two families of bandwidth--latency curves.
Blue curves depict memory-system performance with all prefetchers enabled. This is the default configuration, also used throughout the rest of the study, including in Figures~\ref{fig:actual-hw} and~\ref{fig:gem5-bw-lat-result-characterization-hardware}.
Red curves are measured on the same actual system, but with the L1-D and L2 hardware prefetchers disabled by configuring Model-Specific Registers~(MSRs)~\cite{intel_sdm_vol4}.
With prefetchers disabled, the bandwidth--latency curves cover a smaller portion of the saturated region: 
the maximum memory bandwidths drop by 6--14\,GB/s, while the maximum latencies 
decrease by 74--164\,ns. 
This trend is visible across all read/write compositions of memory traffic. 

\begin{takeawaybox}
\textbf{Finding~8.} 
Memory-system simulation could be further improved by accounting for the time memory requests spend in the memory-controller pipeline and the latency of the memory PHY and I/O. In addition, higher traffic intensities, e.g., those generated by more aggressive data prefetchers, could further increase pressure on the simulated memory systems and reveal additional saturation regions.
\end{takeawaybox}

\section{Required effort}
\label{sec:Effort}

\looseness -1 Although the community recognizes the importance of performance validation for simulation infrastructures, it is also important to consider the effort required to perform such validation. 

For each validation benchmark, we analyze both setup effort and simulation time. 
In our case, the setup was performed by a senior engineer with years of experience using ZSim, Ramulator and DRAMsim3. 
The simulations were executed on a widely available simulation platform: a Latitude 7450 laptop with a 12-core Intel CPU and 16\,GB of DRAM. 
In addition to reporting the absolute setup and simulation times, we compare them with the setup and simulation times of the SPEC CPU2006 suite~\cite{Henning:SPEC} in the same simulation environment. 
The objective is to make this discussion useful for users with different levels of experience and for researchers working with different simulation infrastructures.



Pointer-chase and STREAM are simple and well-known benchmarks, and we consider their setup trivial. 
Mess is more complex than pointer-chase or STREAM, but still much simpler to set up than the full SPEC benchmark suite. 
In our case, the pointer-chase and STREAM setup was completed in less than an hour, Mess required less than a day, and the setup of the full SPEC benchmark suite took less than two days. 

\looseness -1 The simulation time for pointer-chase and STREAM is below 1\,minute.  
A similar amount of time is required to simulate a single Mess data point. 
Our experiments were very detailed, comprising 11 bandwidth--latency curves, each with 40 data points, for a total simulation time of approximately seven hours. 
If needed, the coverage of the experiment can be reduced by considering fewer curves and data points, which proportionally reduces the simulation time. 
For example, simulating five bandwidth--latency curves with 15 data points each requires approximately one hour.   
For comparison, simulating the SPEC CPU2006 suite with 200\,billion instructions per benchmark, which is on the low end of the desired instruction count, required approximately five days.

\begin{takeawaybox}
\textbf{Finding~9.}
\textbf{Pointer-chase} and \textbf{STREAM} should be \textbf{essential sanity checks} for any memory-system simulation. 
If these experiments detect large simulation inaccuracies, or if users require higher confidence in the simulation infrastructure, the \textbf{Mess benchmark} can provide a 
\textbf{detailed and holistic evaluation} of memory-system performance within hours. 
Overall, users have multiple options for evaluating memory-system simulations at moderate cost.
\end{takeawaybox}

\section{Related Work}
\label{sec:related-work}

\subsection{Evaluation methodology}

\looseness -1 Currently, there is no common methodology or set of practices for evaluating memory-simulator performance in integrated simulation infrastructures. As a result, studies adopt different evaluation approaches, which can lead to conflicting conclusions even for the same simulation infrastructure\,\cite{esmaili:mess, bostanci:ramulator-damov-safari}.

For example, the Mess study\,\cite{esmaili:mess} evaluates ZSim integrated with DRAMsim3 and Ramulator and reports significant inaccuracies relative to actual hardware. In contrast, stand-alone, trace-driven DRAMsim3 and Ramulator show substantially higher accuracy, leading the authors to broadly attribute the inaccuracies observed in the integrated simulations to the CPU--memory interfaces. These findings are consistent with the application-view and memory-simulator-view evaluations presented in our work.
 
\looseness -1 The follow-up study by the Ramulator team~\cite{bostanci:ramulator-damov-safari} evaluates the same ZSim+Ramulator infrastructure from a different perspective. The authors argue that DRAM performance should be assessed using statistics reported directly by Ramulator and present results consistent with the trace-driven Ramulator evaluation in the Mess study~\cite{esmaili:mess}. Based on this methodology, they question the application-view evaluation used in the Mess paper, because it is not derived directly from Ramulator outputs. 
 
Our study is the first to explicitly analyze multiple memory-performance views when evaluating memory-simulation accuracy. We show that conclusions about memory-model accuracy in integrated simulation infrastructures must be interpreted in the context of the memory-performance view being evaluated.
The memory-simulator and memory-interface views track memory-system performance across different stages of the integrated simulation infrastructure and are therefore valuable for performance analysis and debugging. The application view directly determines simulated application performance and is therefore the relevant measure of memory-simulation accuracy from the simulator user’s perspective.

Our results also warn that incomplete evaluation methodologies may fail to expose substantial simulation inaccuracies while providing apparent confidence in the accuracy of the overall simulation infrastructure. We hope that this study motivates further community work and discussion in an area that would benefit from additional research and greater methodological rigor.

\subsection{Memory-model incompatibility}
Several previous studies have evaluated parallel system simulators coupled with cycle-accurate memory models and analyzed their compatibility~\cite{bostanci:ramulator-damov-safari, esmaili:mess, Eyerman:IRMM, sanchez:zsimEnhance, Shang:misscycle}. Most of these works address the incompatibility by coupling parallel system simulators with immediate-response memory models~\cite{Eyerman:IRMM, esmaili:mess, Shang:misscycle}. These models can provide accurate memory-performance estimates~\cite{esmaili:mess,crowthers:panem} and represent a suitable choice for many simulation studies.

Our work explores a complementary approach and demonstrates the feasibility of integrating cycle-accurate memory simulators into massively parallel simulation infrastructures. External simulators such as Ramulator, DRAMsim, and DRAMSys are particularly valuable for architectural explorations that require detailed modifications to memory-device timings, protocols, or architectures, including alternative memory technologies and processing-in-memory designs. Such explorations are typically beyond the scope of higher-level immediate-response memory models.

\subsection{Memory-simulator evaluations}
We also review prior work evaluating memory-system performance across a wide range of simulation infrastructures.

\textbf{Memory models in CPU simulators: } 
Only a handful of studies~\cite{sanchez:zsim, Butko:gem5ARM, Gutierrez:gem5ARM} validate the internal memory models of CPU simulators against actual hardware. These evaluations use memory-centric benchmarks, including pointer-chase and \mbox{LMbench~\cite{lmbench}} to measure unloaded memory-access latency, and STREAM~\cite{mccalpin:streamLink, mccalpin:streamBenchmark} to measure sustained memory bandwidth. Extending these studies with the detailed bandwidth--latency characterization proposed in this work would provide a more comprehensive evaluation of memory-system performance. Applying the same methodology to other widely used CPU simulators, such as Sniper~\cite{carlson:sniperOriginal,carlson:sniperUpdated} and ChampSim~\cite{Gober:champsim}, would also be valuable.

\looseness -1 \textbf{gem5 integrated with CXL simulators: }
In 2026, gem5 was integrated with two simulators for CXL memory expanders, both of which use memory-intensive benchmarks to compare the performance of actual and simulated CXL devices. The gem5+CXL-DMSim integration~\cite{Wang:CXLDSim} was evaluated using LMbench~\cite{lmbench}, STREAM~\cite{mccalpin:streamLink, mccalpin:streamBenchmark}, Viper~\cite{Benson:viper}, Redis~\cite{Cooper:Redis}, and MERCI~\cite{Lee:MERCI}. The gem5+Xerxes CXL simulation~\cite{Yuda:Xerxes} provides a quantitative comparison of actual and simulated bandwidth--latency curves.

\textbf{Cycle-accurate memory simulators: }
Since 2011, comparing simulated DRAM timings against manufacturers' Verilog models has been considered the gold standard for memory-simulator validation~\cite{steiner:dramsys4, Rosenfeld:dramsim2, Shangli:dramsim3, kim:ramulator, Haocong:Ramulator2}. However, the recent Mess study~\cite{esmaili:mess} showed that even simulators that do not violate DRAM Verilog timings can still exhibit substantial discrepancies with actual-system performance.

\looseness -1 This finding has influenced recent simulator evaluations, which now also include detailed bandwidth--latency curves~\cite{bostanci:ramulator-damov-safari}. This represents an important improvement over previous evaluation practices. However, these evaluations focus primarily on memory-simulator statistics, often obtained from trace-driven experiments. This is a suitable approach for evaluating the memory simulators themselves, 
but it does not guarantee that applications simulated within complex integrated infrastructures experience the same memory-system behavior.

\section{Conclusions}
\label{sec:Conclusion}

\looseness -1 This paper proposes a methodology that evaluates memory-system performance from three complementary views: the memory simulator, the CPU--memory interface and the simulated application. 
We use these views to detect, understand and correct memory-simulation inaccuracies arising from simulator integration and simplified modeling of microarchitectural features. Most importantly, we identify and correct a fundamental memory-model mismatch between delayed-response cycle-accurate memory simulators and ZSim, a representative parallel CPU simulator.  
This result demonstrates that cycle-accurate memory simulators can be integrated into massively parallel simulation infrastructures.

We also apply our enhancements to ZSim integrations with Ramulator\,2, DRAMsim3, and DRAMSys\,4,
covering all major publicly available cycle-accurate memory simulators. After the corrections and enhancements, all integrations closely follow actual memory-system performance trends:  
STREAM bandwidth differs from actual-hardware measurements by less than 5\% in most experiments, while pointer-chase latency remains 15–18\% lower than on actual hardware.
All simulator enhancements are released as open source.

\section{Acknowledgment}
\label{sec:ack}

\looseness -1 This work was funded by grant PID2023-147979NB-C21 funded by MICIU/AEI /10.13039/501100011033 and by FEDER, EU, 
and grant PID2023-146511NB-I00 funded by MICIU/AEI /10.13039/501100011033 and by FSE+.  
The authors also acknowledge the EuroHPC Virtual Training Academy~(EVITA) project for the HPC education and training framework and platform for hosting curricular materials (Funded by the European Union, Grant agreement No. 101196394).  
We would also like to acknowledge valuable comments from the following individuals: John McCalpin and Pau D\'iaz from the Barcelona Supercomputing Center, Agust\'in Navarro Torres from the University of Zaragoza, and Harald Servat, Mart\'i Torrents, Wim Heirman and Stijn Eyerman from Intel. 
Finally, the authors would like to thank the paper reviewers for their constructive comments, especially the artifact reviewers, who ran thousands of simulation points to confirm the results and quantitative claims presented in the paper.

\appendices
\section{Evaluation of the original DAMOV framework}
\label{sec:damov-evaluation}

Most of the experiments presented in the study  
are based on the integration of ZSim with Ramulator through the interface provided by the DAMOV simulation framework~\cite{Oliveira:damov-ZSim-ramulator}. 
%
To ensure that the inaccuracies we detected are not caused by improper use of the DAMOV interface, we repeat the evaluation on the original DAMOV framework~\cite{Oliveira:damov-ZSim-ramulator}
and detect the same issue\,---\,the application memory-performance view
is effectively decoupled from the memory-simulator statistics.

\textbf{Experimental setup: }
The experimental setup for the evaluation of the original DAMOV framework is summarized in Table~\ref{table:damov-configuration}. We use the default CPU and cache models from DAMOV's \texttt{template\_host\_ooo.cfg} and configure Ramulator with the default DDR4-2400R model.
The memory-bandwidth and latency statistics in the interface and application views are collected using data structures that we added to the simulator's source code.

We also needed to address the incompatibility of the ZSim distribution in the DAMOV repository~\cite{safari:damov-sim} with newer versions of the \texttt{glibc} library. To that end, we applied system-call changes already included in the main ZSim repository~\cite{sanchez:ZSim-repo-pull-req}.  
These minor modifications and environment adjustments do not affect the behavior of the simulator's performance model.
The evaluation charts for the original DAMOV framework can be reproduced using the paper's repository~\cite{esmaili:interfaceGit}.

\begin{table}[!t]
\centering
\caption{Simulation parameters based on DAMOV's
\texttt{template\_host\_ooo.cfg} and the DDR4-2400R memory configuration.}
\label{table:damov-configuration}
\begin{tabular*}{\linewidth}{@{\extracolsep{\fill}}ll}
\toprule
\textbf{ZSim: CPU parameters} & \textbf{Ramulator: Memory parameters} \\
\midrule
Out-of-order core model       & 4$\times$ DDR4-2400R channels          \\
32 cores @ 2.4\,GHz           & 1$\times$ 8\,GB DIMM per channel         \\
~~~Private 32\,KB L1-D cache
                              & 2 ranks per DIMM                       \\
~~~Private 32\,KB L1-I cache
                              & 8 devices per rank                     \\
~~~Private 256\,KB L2 cache
                              & 16 banks per device                    \\
Shared 8\,MB LLC cache         & 32\,MB per device bank                 \\

\bottomrule
\end{tabular*}
\end{table}

\begin{figure}[!t]
  \centering
  \includegraphics[width=.8\linewidth]{graphics/horizontal-legend}
\vspace{-.3cm}

  \subfloat[Memory-simulator view%
           \label{fig:damov-ramulator}]{%
    \includegraphics[width=.5\linewidth]{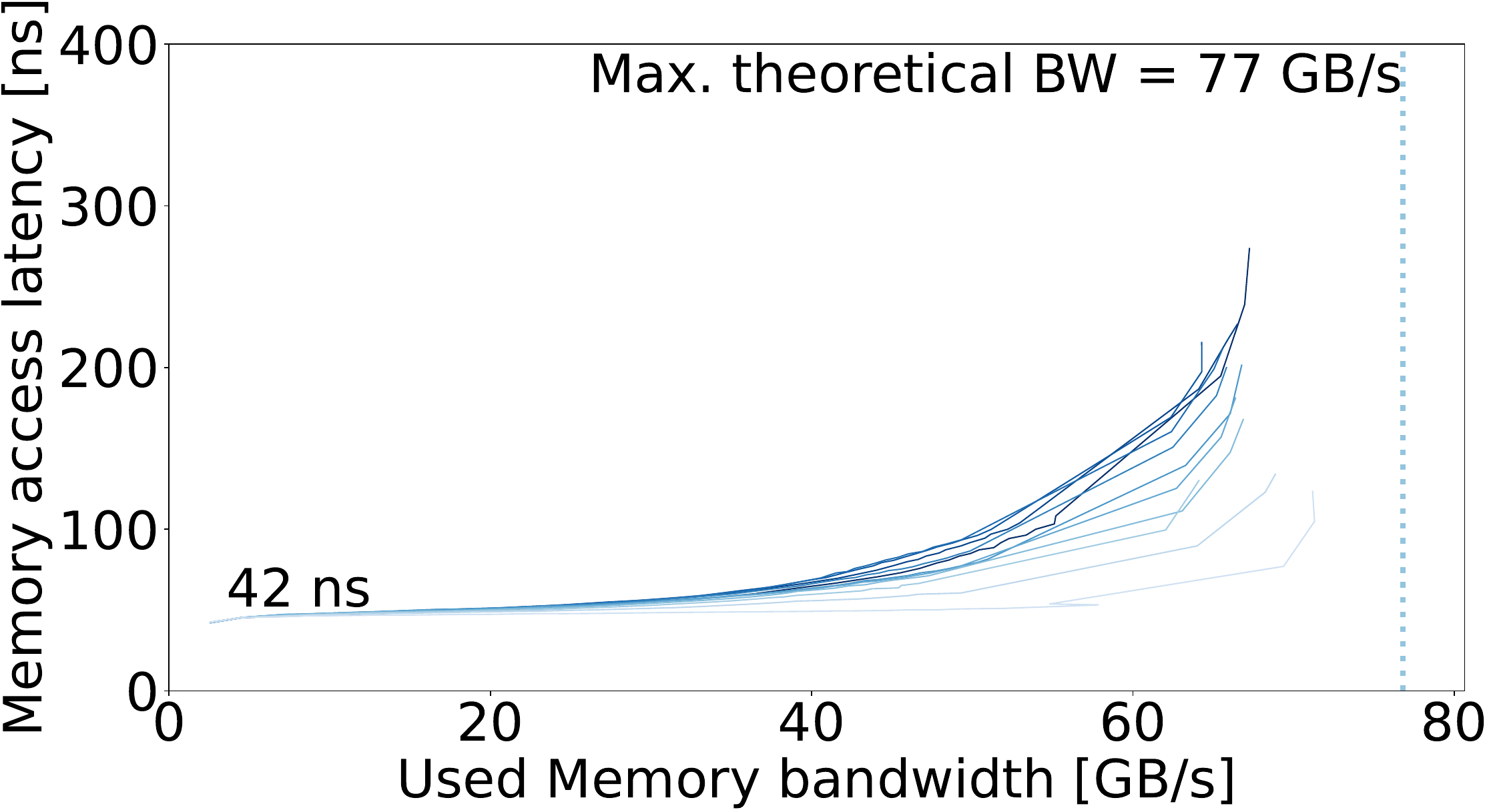}%
  }%
  \subfloat[Memory-interface view%
           \label{fig:damov-interface}]{%
    \includegraphics[width=.5\linewidth]{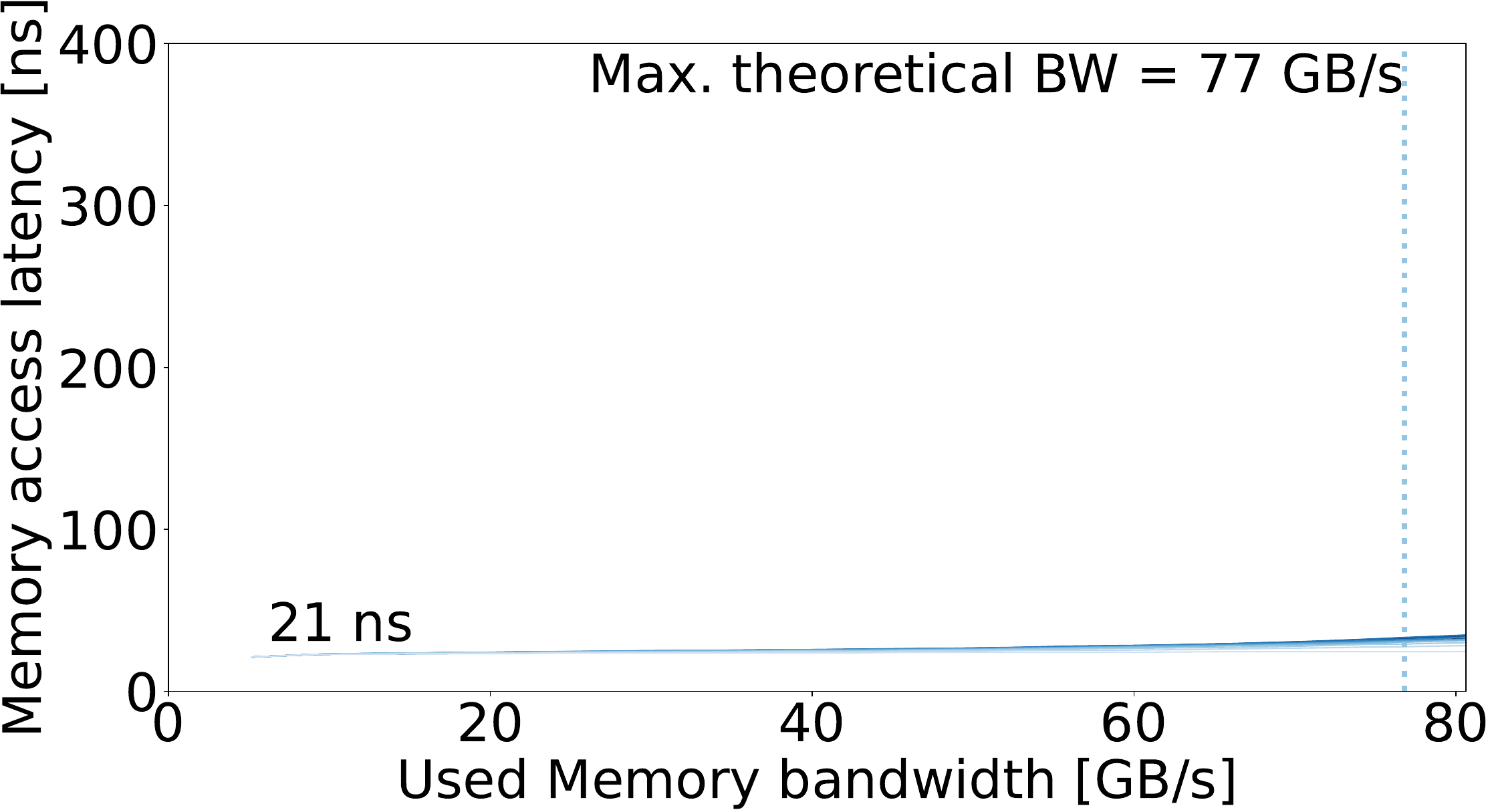}%
  }%
  
  \subfloat[Application view%
           \label{fig:damov-core}]{%
    \includegraphics[width=.5\linewidth]{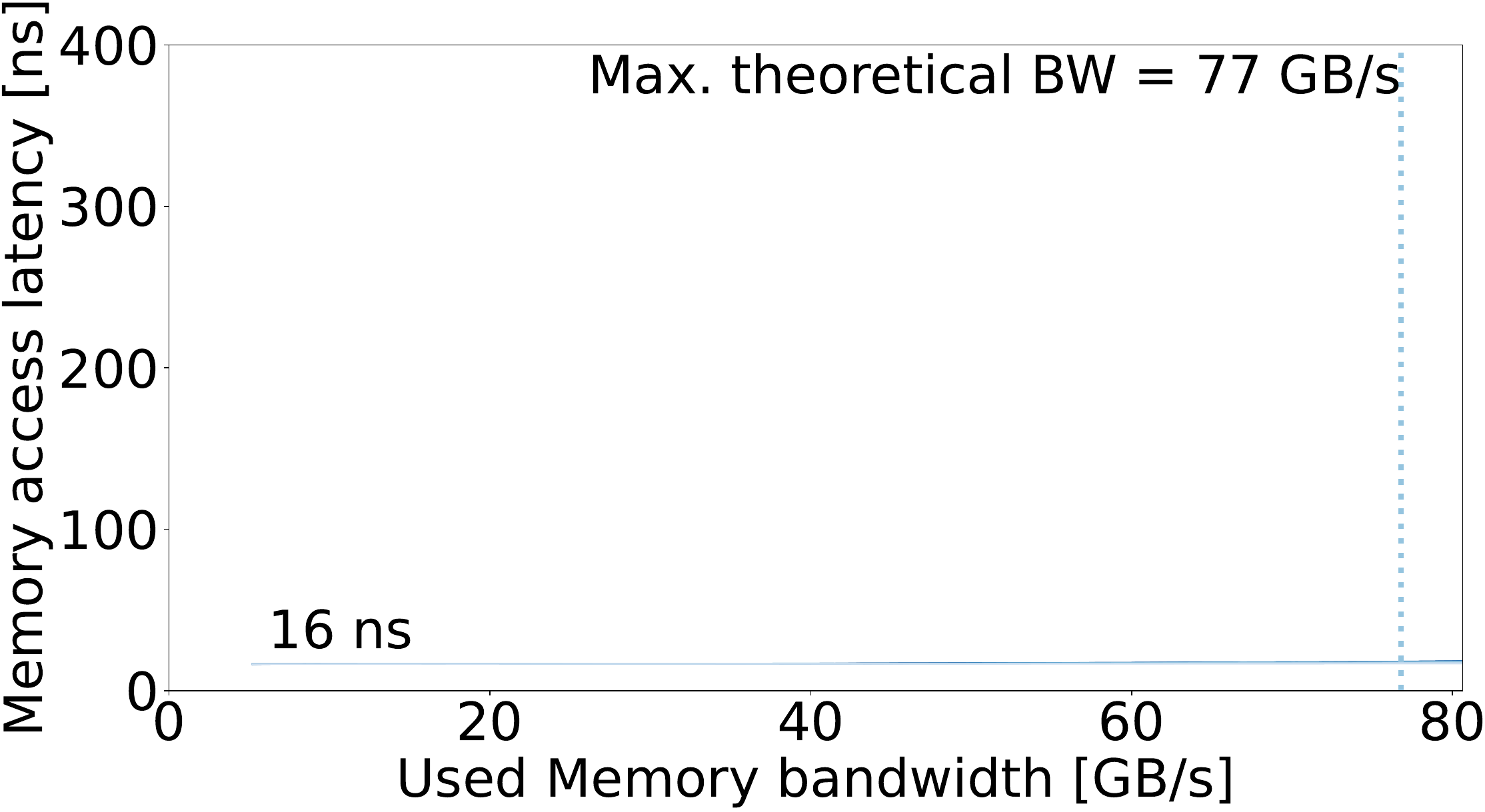}%
  }%
 
  \caption[Four subfigures]{%
  Original DAMOV framework --- The memory-interface and application views of memory-access latency are decoupled from the memory simulator.} 
  \label{fig:original-DAMOV-platform}
\end{figure}

\looseness -1 \textbf{Results and conclusions:} 
%
The results and conclusions of the original DAMOV evaluation (Figure~\ref{fig:original-DAMOV-platform}) closely match our initial evaluation presented in Figure~\ref{fig:original-damov-interface}. 
The memory-simulator view (Figure~\ref{fig:damov-ramulator}), with some 
variations, broadly follows performance trends that could be expected on actual systems. 
The memory latencies in the 
interface view (Figure~\ref{fig:damov-interface})
and the application view (Figure~\ref{fig:damov-core}) 
are unreasonably low: they start at only 21\,ns and 16\,ns, respectively,  
and remain nearly constant across the entire realistic bandwidth range.
Also, the maximum simulated bandwidths in Figures~\ref{fig:damov-interface} and~\ref{fig:damov-core} 
are unrealistically high\,---\,they exceed the theoretical maximum by 36\%. 
Overall, the application’s perception of memory performance is effectively decoupled from the detailed
memory simulation.

\section{Artifact}
\label{sec:ae}
\subsection{Abstract}

This artifact includes the source code and data required to replicate all experiments conducted in our study. We also provide the \texttt{00-damov-native} experiment to demonstrate that the inaccuracies identified in this paper also exist in the original DAMOV platform. This artifact enables readers to understand how the results were obtained, reproduce the experiments on their own systems, compare alternative configurations, and independently evaluate the behavior of the proposed interface beyond our specific platform. 
This artifact is organized as a single code repository plus an experiment data set.
It includes the simulator source trees used in the paper
(\texttt{zsim-bsc}, \texttt{DRAMsim3}, \texttt{DRAMSys4}, \texttt{Ramulator}, and \texttt{Ramulator2}), the benchmarks
(\texttt{ptr\_chase}, \texttt{traffic\_gen}, \texttt{stream-copy}, \texttt{stream-scale},  \texttt{stream-add},  and \texttt{stream-triad}),
experiment-specific configuration files, committed processed outputs, and helper scripts to rerun or compare stages.

The repository organizes each experiment in a separate folder. Each experiment folder
corresponds to one experiment in the paper
and maps directly to one figure or statement. The artifact
covers the progression from the baseline interface
(Figure~2), through the clock-scaling,
corrected-frequency, and memory-model stages (Figures~3,~4,~8), to address
mapping, NOC, and prefetcher refinements
(Figures~9a--9c), the portability experiments
with Ramulator\,2, DRAMsim3, and DRAMSys\,4
(Figures~11b--11d), and finally the memory-intensive benchmarking of the final platform (Figure~11e). For each stage that produces paper figures, the repository stores the processed
bandwidth--latency CSV and the corresponding plots.
Because the raw simulation traces are too large to store
in Git, they are hosted externally. A table in the README
file of each experiment folder provides the download links
for the traces associated with that experiment.


\subsection{Artifact check-list (meta-information)}

{\small
\begin{itemize}
  \item {\bf Program:} ZSim-based CPU--memory simulation platform with Ramulator, Ramulator2, DRAMsim3, and
    DRAMSys backends; pointer-chasing and traffic-generation
    workloads; native DAMOV framework.
  \item {\bf Compilation:} GCC~11 or later (C++20
    required by Ramulator2), \texttt{scons}, \texttt{make},
    CMake 3.25 or later, and Python~3.
  \item {\bf Data set:} Processed and raw simulation traces.  
  \item {\bf Run-time environment:} Linux x86\_64,
    Intel Pin 2.14 (build 71313), HDF5 1.8.16, Python
    with \texttt{pandas} and \texttt{matplotlib}.
  \item {\bf Hardware:} A Linux x86\_64 server or
    workstation. A multicore machine is recommended for
    rerunning full sweeps; inspection of committed outputs
    requires only local file access.
  \item {\bf Metrics:} Memory latency in nanoseconds;
    memory bandwidth in GB/s.
  \item {\bf Output:} Per-stage CSV tables and
    bandwidth--latency curves for the memory simulator
    view, memory interface view, and application view.
  \item {\bf Experiments:} All the experiments in the paper. 
  \item {\bf How much disk space is required
    (approximately)?:} About 0.7\,GB for the checkout;
    allow at least 25\,GB for dependencies, builds, and
    regenerated sweeps.
  \item {\bf How much time is needed to prepare the workflow
    (approximately)?:} 30--60 minutes when Pin and HDF5 are
    already available.
  \item {\bf How much time is needed to complete
    experiments (approximately)?:} The short evaluation (single data point)
    takes minutes; full simulations are stage- and
    hardware-dependent and take substantially longer.
  \item {\bf Publicly available:} Yes, via the artifact repository and externally hosted raw-data packages.

  \item {\bf Code licenses:} BSD 3-Clause for the artifact;
    bundled components retain their upstream licenses.
  \item {\bf Archived (provide DOI)?:} \href{https://doi.org/10.5281/zenodo.21760831}{10.5281/zenodo.21760831}
\end{itemize}
}

\begin{figure}[!t]
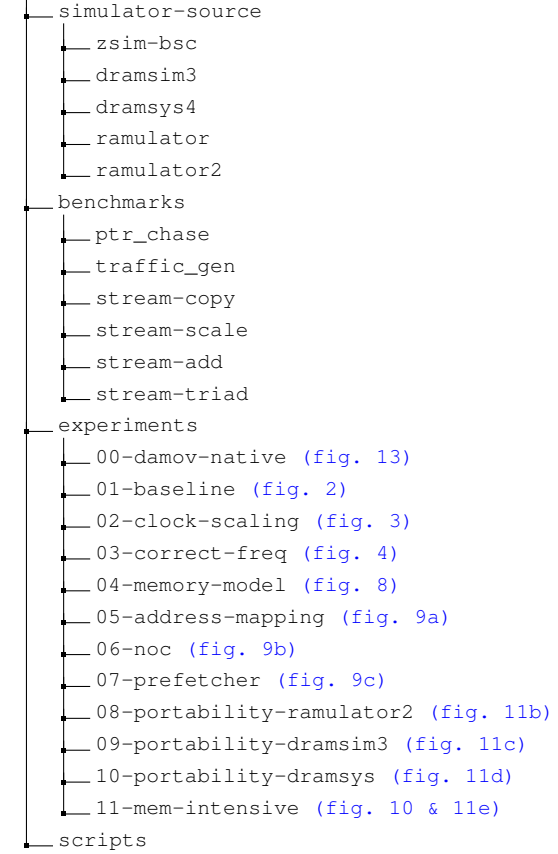

\footnotesize
\dirtree{%
.1 Main GitHub repository.
.2 simulator-source.
.3 zsim-bsc.
.3 dramsim3.
.3 dramsys4.
.3 ramulator.
.3 ramulator2.
.2 benchmarks.
.3 ptr\_chase.
.3 traffic\_gen.
.3 stream-copy.
.3 stream-scale.
.3 stream-add.
.3 stream-triad.
.2 experiments.
.3 00-damov-native \textcolor{blue}{(fig.~13)}.
.3 01-baseline \textcolor{blue}{(fig.~2)}.
.3 02-clock-scaling \textcolor{blue}{(fig.~3)}.
.3 03-correct-freq \textcolor{blue}{(fig.~4)}.
.3 04-memory-model \textcolor{blue}{(fig.~8)}.
.3 05-address-mapping \textcolor{blue}{(fig.~9a)}.
.3 06-noc \textcolor{blue}{(fig.~9b)}.
.3 07-prefetcher \textcolor{blue}{(fig.~9c)}.
.3 08-portability-ramulator2 \textcolor{blue}{(fig.~11b)}.
.3 09-portability-dramsim3 \textcolor{blue}{(fig.~11c)}.
.3 10-portability-dramsys \textcolor{blue}{(fig.~11d)}.
.3 11-mem-intensive \textcolor{blue}{(fig.~10~\&~11e)}.
.2 scripts.
}
\caption{Top-level structure of the artifact repository.}
\label{fig:artifact-tree}
\end{figure}

\subsection{Description}

The artifact is organized around the refinement sequence
presented in the paper. All experiment stages share the same
simulator sources and benchmarks; stages differ only in
their \texttt{sb.cfg} configuration and a small number of
stage-specific overrides. This design makes it possible to
compare intermediate states directly without duplicating the
codebase. 

\subsubsection{How to access}

The artifact is available in the
\href{https://github.com/bsc-mem/cpu-mem-sim-mismatch/tree/main}
{public project repository}.
The default branch is \texttt{main}.  The root
\texttt{README.md} maps each stage to its paper figure and
links the detailed simulator, workload, and experiment
instructions. Raw traces and dependency bundles are hosted on external servers and are linked to the repository.

\subsubsection{Hardware dependencies}

Full simulation targets Linux x86\_64. Intel Pin~2.14 and
\texttt{ptr\_chase} are Linux-specific.  Inspecting
committed results and running the short Python evaluation
does not require the simulator build.  A multicore Linux
machine is recommended for full stage reruns.

\subsubsection{Software dependencies}

See the README inside \texttt{simulator-source/}
for details. The paper used GCC~11.4.0.  Full simulation
requires Intel Pin~2.14 (build~71313), HDF5~1.8.16,
\texttt{scons}, CMake, \texttt{libconfig++}, and Python~3
with \texttt{pandas} and \texttt{matplotlib}. If Pin and HDF5 are not detected on the host system, they will be automatically downloaded by \texttt{scripts/setup-env.sh}.

\subsubsection{Data sets}

For each experiment in the paper, the repository
commits: 
\begin{itemize}
  \item The  configuration (\texttt{sb.cfg} and
    optional overrides),
  \item The processed CSV
    (\texttt{bandwidth\_latency.csv}),
  \item The generated plots under \texttt{figures/}, and
  \item The link in README that points to the
    externally hosted raw simulation traces.
\end{itemize}

The committed figures use the same terminology as the paper:
memory-simulator view, memory-interface view, and
application view. The portability stages provide the 
memory-interface and application views; the memory-simulator view
is omitted when the corresponding data is unavailable.

\subsection{Installation}

Clone the project using the repository
link above and run the one-shot setup:

\begin{lstlisting}[language=sh, frame=single,
    basicstyle=\scriptsize, columns=fullflexible]
git clone https://github.com/bsc-mem/cpu-mem-sim-mismatch.git
cd cpu-mem-sim-mismatch
./setup.sh
source .zsim-env
\end{lstlisting}

The script checks dependencies, downloads external packages and builds the simulator backends, ZSim, and shared workloads.

\subsection{Experiment workflow}

Once the project is built, run an experiment through the shared runner (\texttt{experiments/runner.sh}). The runner produces raw simulation data. The following commands run the \texttt{01-baseline} experiment, which corresponds to Figures~2b, 2c, and 2d:

\begin{lstlisting}[language=sh, frame=single,
    basicstyle=\scriptsize, columns=fullflexible]
source .zsim-env
./experiments/runner.sh 01-baseline
\end{lstlisting}

 Raw output is in \texttt{experiments/01-baseline/test-raw}. Pass this directory and the experiment configuration directory to the shared plotting script (\texttt{experiments/plot.py}) to generate the corresponding figures:
 
\begin{lstlisting}[language=sh, frame=single,
    basicstyle=\scriptsize, columns=fullflexible]
./experiments/plot.py \
  experiments/01-baseline/test-raw \
  --config-dir experiments/01-baseline
\end{lstlisting}

The plotting script writes the processed CSV file and the three views corresponding to Figures~2b, 2c, and 2d under:

\begin{lstlisting}[language=sh, frame=single,
    basicstyle=\scriptsize, columns=fullflexible]
test-output/01-baseline/processed/
test-output/01-baseline/figures/
\end{lstlisting}

To reproduce another experiment from 01 through 10, replace \texttt{01-baseline} in the runner command and both paths passed to the plotting script.

Experiment 11 uses a different raw-data layout and has its own plotting script. The following commands run Experiment 11 and generate Figures~10 and 11e:

\begin{lstlisting}[language=sh, frame=single,
    basicstyle=\scriptsize, columns=fullflexible]
./experiments/runner.sh 11-mem-intensive
./experiments/11-mem-intensive/plot.py
\end{lstlisting}

Experiment 11 writes its output figures to the directory address \texttt{experiments/11-mem-intensive/figures/} and its processed CSV file to \texttt{experiments/11-mem-intensive/processed/}.

\subsection{Evaluation and Expected Results}

Running experiments 01 through 10 generates the bandwidth--latency curves from the paper and saves them to the default output directory: \texttt{test-output/<stage>/figures/}. For Experiment 11, system and runtime variation may cause differences of approximately 1--4\% in individual benchmark results. These differences do not affect the paper's trends or conclusions. Exact reproduction of the published figures requires the raw data available in the repository or through the provided links. The following commands download the raw results for Experiment 01 and regenerate Figures~2b,~2c, and~2d:

\begin{lstlisting}[language=sh, frame=single,
    basicstyle=\scriptsize, columns=fullflexible]
curl -fL -o /tmp/01-baseline.zip \
  'https://zenodo.org/records/21760832/files/01-baseline.zip?download=1'

unzip -o /tmp/01-baseline.zip -d experiments/01-baseline

./experiments/plot.py \
  experiments/01-baseline/test-raw \
  --config-dir experiments/01-baseline 
\end{lstlisting}

The plotting script stores the generated figures under \texttt{test-output/01-baseline/figures/}. Experiments 02 through 10 follow the same steps. Their raw data is archived externally, with download links provided in the repository.







\bibliographystyle{IEEEtran}
\bibliography{ref}

\end{document}